\documentclass[3p]{elsarticle}

\usepackage{afterpage}
\usepackage{algorithm}
\usepackage{algpseudocode}
\usepackage{rotating}
\usepackage{lineno,hyperref}
\usepackage{graphics}
\usepackage{arabtex}
\usepackage{utf8}
\usepackage[table]{xcolor}
\usepackage{multirow}
\usepackage{makecell}
\usepackage{longtable}
\usepackage{pdfpages}
\usepackage{pdflscape}
\usepackage{tabularx}
\usepackage{amsmath}
\usepackage{amssymb}
\usepackage{arydshln}
\usepackage{array}
\def\endabstract{\egroup}

\journal{Journal of Ambient Intelligence and Humanized Computing}

\biboptions{authoryear}

\begin{document}

\begin{frontmatter}

\title{A Model to Measure the Spread Power of Rumors}


\author[address1,address2,address8]{Zoleikha Jahanbakhsh-Nagadeh}
\ead{zoleikha.jahanbakhsh@iau.ac.ir}

\author[address1]{Mohammad-Reza Feizi-Derakhshi\corref{mycorrespondingauthor}}
\cortext[mycorrespondingauthor]{This is to indicate the corresponding author.}
\ead{mfeizi@tabrizu.ac.ir}

\author[address1]{Majid Ramezani}
\ead{m_ramezani@tabrizu.ac.ir}

\author[address1,address3,address7]{Taymaz Akan (Rahkar Farshi)}
\ead{taymazfarshi@ayvansaray.edu.tr}

\author[address1]{Meysam Asgari-Chenaghlu}
\ead{m.asgari@tabrizu.ac.ir}

\author[address1]{Narjes Nikzad–Khasmakhi}
\ead{n.nikzad@tabrizu.ac.ir}

\author[address1]{Ali-Reza Feizi-Derakhshi}
\ead{derakhshi96@ms.tabrizu.ac.ir}

\author[address1,address4]{Mehrdad Ranjbar-Khadivi}
\ead{mehrdad.khadivi@iaushab.ac.ir}

\author[address1,address5]{Elnaz Zafarani-Moattar}
\ead{e.zafarani@iaut.ac.ir}

\author[address6]{Mohammad-Ali Balafar}
\ead{balafarila@tabrizu.ac.ir}

\address[address1]{Computerized Intelligence Systems Laboratory, Department of Computer Engineering, University of Tabriz, Tabriz, Iran.}
\address[address2]{Department of Computer Engineering, Naghadeh Branch, Islamic Azad University, Naghadeh, Iran.}
\address[address8]{Department of Computer, Science and Research Branch, Islamic Azad University, Tehran, Iran}
\address[address3]{Department of Software Engineering, Istanbul Topkapi University, Istanbul, Turkey.}
\address[address7]{Clinical Informatics, Louisiana State University Health Sciences Center Shreveport, Shreveport, USA.}
\address[address4]{Department of Computer Engineering, Shabestar Branch, Islamic Azad University, Shabestar, Iran.}
\address[address5]{Department of Computer Engineering, Tabriz Branch, Islamic Azad University, Tabriz, Iran.}
\address[address6]{Department of Computer Engineering, University of Tabriz, Iran.}

\begin{abstract}
With technologies that have democratized the production and reproduction of information, a significant portion of daily interacted posts in social media has been infected by rumors. Despite the extensive research on rumor detection and verification, so far, the problem of calculating the spread power of rumors has not been considered. To address this research gap, the present study seeks a model to calculate the Spread Power of Rumor (SPR) as the function of content-based features in two categories: False Rumor (FR) and True Rumor (TR). For this purpose, the theory of Allport and Postman will be adopted, which it claims that importance and ambiguity are the key variables in rumor-mongering and the power of rumor. Totally 42 content features in two categories "importance" (28 features) and "ambiguity" (14 features) are introduced to compute SPR. The proposed model is evaluated on two datasets, Twitter and Telegram. The results showed that (i) the spread power of False Rumor documents is rarely more than True Rumors. (ii) there is a significant difference between the SPR means of two groups False Rumor and True Rumor. (iii) SPR as a criterion can have a positive impact on distinguishing False Rumors and True Rumors.
\end{abstract}


\end{frontmatter}


\section{Introduction}\label{Sec:introduction}

\begin{center}
	\textit{“You can get a face mask exemption card so you don’t need to wear a mask”.}
	\newline
	\textit{“A vaccine to cure COVID-19 is available”.}
	\newline
	\textit{“The new coronavirus was deliberately created or released by people”.}
\end{center}

Nowadays, increasing numbers of people join OSNs for daily communications or even business activities \cite{jiang2016understanding}. Many people exchange different messages through messengers and social media, unaware of their factual accuracy. A significant portion of these messages is fake news or rumors, which have an undeniable impact on different aspects of our life. DiFonzo and Bordia \cite{DiFonzo2007} have defined rumor as unverified and instrumentally relevant information statements in circulation that arise in contexts of ambiguity, danger or potential threat, and that functions to help people make sense and manage risk". This unverified information may turn out to be true, or partly or entirely false; alternatively, it may also remain unresolved. Accordingly, we classified rumors into two categories: False Rumor (FR) and True Rumor (TR). An FR is misinformation or inaccurate information that is designed for certain goals based on special content features, while TR is a real social fact. Also, the users who create and spread rumors is named as rumormongers.

Many rumors are spread depending on the social, economic, political and cultural conditions that can appear as one of the causes of anxiety in society and cause frustration among people in society. To clarify the role of spreading rumors in creating insecurity and disturbing the public mind, we can refer to the rumors published in the 2016 U.S. presidential election. In that year, various rumors were spreading on social media (Twitter and Facebook) during the election, so that, among all the 1,723 checked rumors from the popular rumor debunking website Snopes.com, 303 rumors were about Donald Trump and 226 rumors were about Hillary Clinton. Thereby, these rumors could potentially have negative impacts on their campaigns \cite{Jin2017}. 

Rumor is an important and big problem because it can have high destructive power in society. Regardless of the validity of this information, the spread of information is faster than ever. This brings unprecedented challenges in ensuring the reliability of the information \cite{ALZANIN2018294}. This is so important that it has been noticed by the largest IT companies such as Twitter, Facebook and Google. They try to validate the messages during their publication and display the validation result to the user. Therefore, it is very important to identify the rumor in the early hours of the release and to prevent its harmful consequences.

Many researchers have analyzed the problem of rumor in different fields, including detection, validation, stance detection, rumor source detection, propagation struct modeling, and so on. They have used various features at different levels of content, user, and propagation network. In this study, the problem of the rumor is studied in a different field from other research. We have introduced a new index based on content features to calculate the spread power of rumors. Rumor is a collective effort that uses the power of words to interpret a vague but fascinating situation, so we hypothesized that the first influential factor in spreading a rumor is the content of the rumor. In other words, it is the power of words that empowers the text to influence the audience. Because, in the initial moments of spreading a rumor, there is not enough information about the users and the structure of the rumor, but only the content information of the rumor can be explored. Thereby, we proposed a model to calculate the spread power of rumor for the first time and named as the Spread Power of Rumor (SPR). This model is based on content features - at the semantic, syntactic, practical, and lexical levels - of message documents.

To the best of our knowledge, there is no work in the area of SPRs, except Allport and Postman’s theory \cite{Allport1947} about the power of rumors. Allport and Postman presented a hypothesis based on rumor psychology and discussed the power of spread from a psychological perspective. According to their hypothesis, there are two basic conditions for spreading a rumor: (1) the importance of the subject of the rumor for the audience, (2) the existence of ambiguity in the expression of the subject. They defined the law of rumor based on the product of importance and ambiguity. In this study, this psychology hypothesis is mathematically modeled for the first time.

The importance of calculating the spread power of a message is due to the importance of early detection of rumors. The resources available at the beginning of the rumor propagation are so limited, so early rumor detection in the early hours is very challenging. The first factor influencing the message in the early hours of its propagation is its content characteristics. Therefore, introducing and using more new content features can be effective in the early detection of rumors. To this end, we proposed the SPR score as a content-based feature and independent of time-based features. We hypothesize that the SPR criterion can distinguish False Rumor (FR) from True Rumor (TR). To prove this hypothesis, we investigated the efficiency of the SPR criterion in the problem of classifying rumors into two classes FR and TR. 

Allport and Postman proposed the theory of the power of rumor by analyzing the psychological issues of English rumors. Hence, this problem can be implemented by analyzing the content of rumors in different languages. Persian language is used in this study. One of the main motivations about using Persian language is because of low resources of it. Solving this problem in a low resource language shows that proposed approach can be also used for other languages as well. Another motivation is that the authors are native in the Persian language and can analyze dependencies between proposed features and results much better.Therefore, the content characteristics of Persian rumors are analyzed at different lexical, syntactic, and semantic levels. 

The contribution of this research is as follows:

\begin{itemize}
	\item Computing SPR for the first time. The main purpose of this study is to present a mathematical measure called Spread Power of Rumor (SPR) as a new work in the field of rumor analysis in online media. SPR calculation is a research issue that has not been investigated on rumors in any previous studies empirically. This research is the first work that mathematically formulated SPR. The spread power of a rumor depends on two factors: the importance and the ambiguity. We have introduced and evaluated a set of content features which can be used to measure importance and ambiguity of a text.
	\item Investigating significance of SPR. We want to introduce SPR as a criterion in rumor detection task. Therefore, a T-test has been performed to prove significant difference of SPR between FR and TR. The results of this test show that there is a significant difference between the spread power of the two classes TR and FR so that the spread power of FRs is more than TR. Therefore, SPR can be used as a feature in the rumor detection process. 
	\item Application of SPR in rumor detection. We intend to demonstrate one of the applications of SPR in rumor analysis. Therefore, we used SPR in detecting rumors. Rumors are categorized based on a set of content-based features, once without considering the SPR criterion and again with the SPR. The results shown that the presence of SPR as a diagnostic criterion can be effective in categorizing rumors.
\end{itemize}

The rest of the paper is organized as follows: In Section \ref{Sec:theory}, theory of the law of the rumor is defined.  In Section \ref{Sec:Problem}, the problem definition and objectives are presented. Section \ref{Sec:Related works} reviews a summary of related works. Section \ref{Sec:Methodology} describes the proposed SPR measurement model and investigates a set of effective features to compute SPR. Section \ref{Sec:EXPERIMENTS AND RESULTS} describes the experiments and evaluations and in Section \ref{Sec:Conclusion} discussion and conclusions of the paper is shown. Finally in section \ref{Sec:future_work} suggestions for future research are represented.

\section{Baseline Theory}\label{Sec:theory}
Every social phenomenon needs a special set of conditions for its emergence and reliability. Rumor is also a social phenomenon that requires conditions for publication and acceptance. Now, the question is, what conditions and factors are needed to spread the rumor? To answer this question, \cite{Allport1947} outlined two fundamental conditions for spreading the rumor: first, the issue of rumor should be important to the audience. If the subject is interesting to the audience, rumors about that subject may be interesting to them, but this condition alone is not enough. The second condition for propagating the rumor is the existence of ambiguity in expressing the issue. Of course, rumors are more infectious; when little information is released through authoritative channels and uncertainties occur in society. Thereby, \cite{Allport1947} defined the law of the power of the rumor based on the multiplication of importance and ambiguity. So far, this law has been presented as a theory and has not been practically studied on rumors.

\begin{equation} \label{eq:allport}
	Power \approx Importance \times Ambiguity
\end{equation}

In formula \ref{eq:allport}, the relation between "Importance" and "Ambiguity" is not sum, but is multiplication; because if ambiguity or importance is zero then there will be no rumor. According to this theory, whenever the importance of a rumor is high, its influence rate goes up equally. Also, with increasing ambiguity in the case, the penetration rate of rumors rises. If one of these factors be zero, the influence rate of the rumor will be zero \cite{Allport1947}. In other words, it is unlikely that an individual will attempt to spread a rumor that does not matter to him, although it is ambiguous. Also, the importance of the subject alone is not enough to spread the rumor, because the importance should be with the ambiguity that the rumor reveals. For example, a rumor about choosing a presidential candidate after the announcement of the election results, though it is an important issue, due to the lack of ambiguity, it will not spread.

As another inference, a rumor about raising or lowering the percentage of banks' profits does not matter to anyone who does not have the money in the banks, and he will not pursue it. On the other hand, such rumors are less effective for bank employees and officials as they are aware of the exact news of the profit levels in the banks and there is no ambiguity for them, but it is different for ordinary people.

This theory is based on two assumptions \cite{Allport1947}: (i) people exert effort to find meaning in things and events; (ii) when people faced with ambiguity in any important matter, they try to find some meaning by retelling related rumors. This means that the importance and ambiguity of rumors are vital variables that predict whether a rumor would be transmitted or not \cite{Allport1947}.

\section{Problem Statement}\label{Sec:Problem}

The problem of rumor detection is considered as a classification problem that it usually is either a binary (true or false) or a multi-class (true, false or unverified) classification problem \cite{li2019rumor}. Classification of text documents involves assigning a text document to a set of pre-defined classes. Let $D = \{d_1, d_2, …, d_n\}$ be the set of $n$ rumor document which is in two classes FR and TR. $d \in D$ is a document that contains a sequence of m sentences (i.e., $d = S_1, S_2, …, S_m$). $S_i(d)$ is $i$th sentence from document $d$.  Each sentence $S$ has a sequence of k tokens (i.e., $S = t_1, t_2, …, t_k$), including terms, punctuations, numbers, and symbols. Since all message documents $d$ are either TR or FR, it can be inferred that:

\begin{equation} \label{eq:2_2}
	P(FR | d) = 1 - P(TR | d)
\end{equation}

Each rumor has a degree of spread power. Based on our hypothesis about the difference in spread power of FRs and TRs, it can be argued that the SPR criterion has a decisive role in determining $P(FR | d)$. So that $P(FR | d)$ and SPR have a proportional ratio (equation \ref{eq:2_3}):

\begin{equation} \label{eq:2_3}
P(FR | d) \propto SPR
\end{equation}

Now, according to this assumption and the theory of Allport and Postman \cite{Allport1947}, the first question which arose is that:

\begin{itemize}
	\item \textbf{Q1:} How the spread power of a document will be calculated?
\end{itemize}

According to Allport and Postman's theory, SPR is approximately equal to the multiplication of the importance and ambiguity  surrounding the rumor.

We used formula \ref{eq:allport} as a principle and proposed equation \ref{eq:2} to \ref{eq:2_4} in computing the SPR based on content information of rumors. Therefore, the present research seek to represent and compute two factors \textit{"Importance" (Imp)} and \textit{"Ambiguity" (Amb)} of the rumor based on its content information. \newline Therefore, it is necessary  need to address the problem in more detail. Thereby, two other questions arise:

\begin{itemize}
	\item \textbf{Q2:} What content features show importance of the rumor?
	\item \textbf{Q3:} What content features show ambiguity of the rumor?
\end{itemize}

This study analyzed the content features of the rumors to answer these questions. We have shown that analyzing content-based features using Natural Language Processing (NLP) methods can provide useful information, because, the power of the words used in the document should not be underestimated.

In the following, we seek to assess SPR as a function of content-based features between FRs and TRs, and investigate the role of SPR in verifying rumors. Therefore, Q 1 to 3 questions are followed by another question based on which,

\begin{itemize}
	\item \textbf{Q4:} Is there a difference between the spread power of False-Rumors and True-Rumors?
\end{itemize}

To answer this question, it is necessary to compute the spread power for a large set of TRs and FRs separately and then perform evaluations such as t-test on them to determine whether the SPR criterion is effective in distinguishing FRs from TRs. This study seeks to address these questions and aims at investigating the effect of SPR on verifying the rumors.

\section{Related works}\label{Sec:Related works}
Rumors have been extensively studied in the fields of psychology, sociology, and epidemics for decades. Also, the problem of detecting and verifying rumors has been considered in recent years, and continuous progress has been made in this regard. Various approaches have been proposed for analyzing rumors in previous research.

In the area of psychology, as mentioned, \cite{Allport1947} stated that the rumors are the product of two factors, "importance" and "ambiguity", which are two determining factors in the power of rumors. \cite{Harsin2006} presented the idea of the "Rumor Bomb", which means that a "Rumor Bomb" spreads the notion of the rumor into a political communication concept. For Harsin, a "rumor bomb" extends the definition of rumor into a political communication concept with the following features: (i) a crisis of verification. (ii) a context of public uncertainty or anxiety about a political group, figure, or cause, which the rumor bomb overcomes or transfers onto an opponent. (iii) a partisan, which seeks to profit politically from the rumor bomb’s diffusion. (iv) a rapid diffusion via social media. In another research, \cite{Kumar2014} explored the use of theories in cognitive psychology and proposed an algorithm that would use social media as a filter to separate misinformation from accurate information. The cognitive process involved in the decision to spread information involves answering four main questions viz consistency of message, the coherency of the message, the credibility of the source, and general acceptability of message. They proposed an algorithm that uses the collaborative filtering property of social networks to measure the credibility of sources of information as well as the quality of news items.

 In this study, \cite{Allport1947} 's theory about the law of the rumor is mathematically modeled and presented as a criterion for calculating the spread power of rumors. It is necessary to examine SPR on an application of rumors to evaluate the effectiveness of the SPR criterion in distinguishing rumors from non-rumors. To this end, we evaluated the SPR on the problem of rumor detection. To automatically solve the rumor detection problem, various researches are presented with different modeling methods, detecting, verifying, and preventing rumors by analyzing various features at three levels: user information, content, and propagation network structure on many languages \cite{ZHANG2019102025}. Some of them are classical learning methods like Naive Bayes (NB), Support Vector Machines (SVM), Decision Tree (DT), Random Forest (RF), Logistic Regression (LR), and others are based on deep learning methods \cite{8005992}. Most classical machine learning models follow the popular two-step procedure, wherein the first step some hand-crafted features are extracted from the documents (or any other textual unit) and in the second step those features are fed to a classifier to make a prediction \cite{minaee2020deep}. Since this study utilized the content features extracted by feature engineering methods to calculate the spread power of a message document, so we discussed the machine learning-based classification systems and skipped other methods such as methods based on deep neural networks. Table \ref{tbl:rel-work} presents a brief literature review of veracity and detection classification using the machine learning methods.

In classical machine learning approaches, researchers analyzed various features at three levels: user information, content, and propagation network structure \cite{ZHANG2019102025}. For example, \cite{Castillo2011} and \cite{Kwon2013a} proposed a combination of linguistics and structure-based features that can be used to approximate the credibility of information on Twitter \cite{Castillo2011} studied the propagation of rumors during real-world emergencies while \cite{Kwon2013a} studied the propagation of urban legends (such as Bigfoot) on Twitter.  \cite{Yang2012} have done work similar to Castillo’s work on Sina Weibo (a Chinese leading micro-blogging service provider that functions like a FacebookTwitter hybrid). \cite{Kwon2013a} and \cite{Yang2012} show that the most significant features for rumor detection are emoticons, opinion words, and sentiment scores (positive or negative). \cite{Qazvinian2011} explored the content-based, network-based features, and microblog-specific memes to address the problem of rumor detection. \cite{Wu2015} used all previous effective features, plus two new semantic features: a topic model feature and a search engine feature. \cite{Vosoughi2015} identifies salient features of rumors in different periods by analyzing three aspects of information spread: linguistic, user, and network propagation dynamics using Dynamic time wrapping (DTW) and Hidden Markov Models(HMMs). \cite{Wang2015} identified a series of short diffusion patterns, based on stance, that appear to be strongly related to rumors. \cite{floos_2016} presented a statistical method based on the computation of TF-IDF for each term in the tweets to detect rumors in Arabic tweets. \cite{Zhao:2015:EME:2736277.2741637} tackled early detection of rumors by determining clusters of potential rumors and extracted a series of features for each cluster. These features were two types of language patterns in rumors: the correction type and the inquiry type. \cite{Liu2016} proposed a model based on propagation patterns of rumors and credible messages and carried this model on Sina Weibo. \cite{zubiaga2017exploiting} leveraged the context preceding a tweet with a sequential that learns the reporting dynamics during an event to detect rumors. Their proposed method was based on the hypothesis that a tweet alone may not suffice to know if its underlying story is a rumor, due to the lack of context. Kwon et al. \cite{Kwon2017} examined user, linguistic, network, and temporal features over different observation time windows. They identified significant differences between rumors and non-rumors for the first 3, 7, 14, 28, and 56 days from the initiation). Zamani et al. \cite{Zamani2017} addressed the problem of rumor detection on Persian Twitter for the first time and developed a dataset of Persian Twitter rumors. They utilized a set of structural features based on tweet and user characteristics, and also used frequent Twitter unigrams as words vector. Zarharan et al. \cite{zarharanpersian} focused on the stance detection of Persian rumor and developed a dataset for it. Jahanbakhsh et al. \cite{jahanbakhsh2020speech} proposed a dictionary-based statistical technique to identify Persian SA. Therefore, Based on the obtained results in \cite{jahanbakhsh2020speech}, FRs are often expressed in four SA classes, including threat (SA Thrt), declaration (SA Dec), question (SA Ques), request (SA Req). They showed the positive effect of SA on rumor detection by combining the content features (into four categories: Lexical, Semantic, Syntactic, and Surface) and four speech act classes. Kumar et al. \cite{Kumar2019} first extracted three categories of features, including content-based features (Part-of-Speech, Bag of Words, term-frequency), pragmatic features (emoticons, sentiment, anxiety-related words and Named Entity), and network-specific features (User and Message metadata). Then, they used particle swarm optimization to select the set of features with the highest importance on the rumor veracity classification task. 

Some research has also focused on modeling the distribution of rumors in the social space \cite{ZHANG2019102025}. For example, \cite{Zeng2016} modeled the speed of information transmission to compare retransmission times across content and context features. Doerr et al. \cite{Doer2012} simulated a natural rumor spreading process on several classical network topologies. They also performed a mathematical analysis of this process in preferential attachment graphs and proved that the process of rumor spreading disseminates a piece of news in sub-logarithmic time. That is, the spread of the rumors is extremely fast on social networks. Based on these results, it can be argued that the spread power of FRs is more than TRs.

In existing research, the problem of rumor has been discussed from various fields, including theories of rumor psychology, automatic diagnosis and validation, and modeling of rumor propagation. Each of these works introduces features at different levels of content, user, and propagation network. The information that can be extracted from the rumor published in the early hours of publication is limited. Therefore, the features introduced in previous research are more dependent on information about the history of rumor spread, such as user information and network structure, and they are less focused on content features. Therefore, the present study intends to introduce a criterion as a feature that can be used in the early detection of rumors. To date, the Allport-Postman Rumor Law \cite{Allport1947} (i.e., the power of rumor) has not been mathematically established in any research. To this end, this study intends to model the rumor law mathematically for the first time and calculate the Spread Power of Rumor (SPR) based on its content characteristics and without dependence on the background information of the rumor. We also prove that there is a significant difference in SPR between FR and TR. Hence, SPR can be used as a new and effective criterion in distinguishing between rumor and non-rumor in the early moments of publication.

\begin{table}
	\caption{The list of previous machine learning techniques for rumor detection based on Content (C), User (U), and Structural (S) features.}
	\label{tbl:rel-work}
	\begin{tabular}{m{1cm} m{0.7cm} m{1.9cm} m{3.3cm} m{0.2cm} m{0.2cm} m{0.2cm} m{5.5cm}}
		& 	& 	& 	& \multicolumn{3}{c}{\textbf{Features}} &  \\
		\multirow{-2}{*}{\textbf{Ref.}}	&\multirow{-2}{*}{\textbf{Lang.}}	&\multirow{-2}{*}{\textbf{Dataset}}	&\multirow{-2}{*}{\textbf{Method}}	&\textbf{C}	&\textbf{U}	&\textbf{S}	&\multirow{-2}{*}{\textbf{Conclusion}} \\
		\hline
		\cite{Castillo2011}	&EN	 &Twitter &DT, NB, SVM. &\checkmark &\checkmark &\checkmark &DT as best classifier. \\
		\rowcolor{gray!15} \cite{Qazvinian2011}	&EN	&Twitter &Present the tweet with two patterns: Lexical and Part-of-speech.	&\checkmark	&\checkmark	&\checkmark	&Identify users that spread false information in online social media using their proposed features.	\\
		\cite{Yang2012}	&CHI	&Sina Weibo	&SVM-RBF kernel &\checkmark &\checkmark &\checkmark &Improve in accuracy. \\
		\rowcolor{gray!15} \cite{Kwon2013a}	&EN	&Twitter &DT, RF, SVM, LR &\checkmark &  &\checkmark &RF as best classifier. \\
		\cite{Wu2015}	&CHI	&Sina Weibo &SVM-RBF kernel &  &  &\checkmark &Improve in accuracy using network based features. \\
		\rowcolor{gray!15} \cite{Wang2015}	&EN	&Twitter &Analyze patterns of diffusion with linear model &  &  &\checkmark &Identify influential spreaders. \\
		\cite{Vosoughi2015}	&EN	&Twitter &Verify rumors in different time periods using DTW and HMMs &\checkmark &\checkmark &\checkmark &HMM as best classifier. \\
		\rowcolor{gray!15} \cite{floos_2016}	&AR	&Twitter &TF-IDF &\checkmark &  &  &The effectiveness of content features in validating Arabic tweets \\
		\cite{Zhao:2015:EME:2736277.2741637}	&En	&Twitter &Searching enquiry phrases, clustering similar posts, then ranking the clusters. &\checkmark  &  &  &Accuracy of 0.52 for their best run using J48. \\
		\rowcolor{gray!15} \cite{Liu2016}	&CHI	&Sina Weibo &SVM &  &\checkmark &  &Differences in the propagation patterns of rumors and credible messages \\
		\cite{zubiaga2017exploiting}	&En	&Twitter &A sequential classifier &  &\checkmark &\checkmark &  \\
		\rowcolor{gray!15} \cite{Kwon2017}	&EN	&Twitter &RF &\checkmark &\checkmark &\checkmark &Identify significant features in the first 3, 7, 14, 28 and 56 days of the initiation. \\
		\cite{Zamani2017}	&FA	&Twitter &J48, Naive Bayes, SMO, IBK &\checkmark &\checkmark  &\checkmark &About 70\% precision just based on structural features and about 80\% based on both categories of features. \\
		\rowcolor{gray!15} \cite{Mahmoodabad2018}	&FA	&Twitter &MLP, KNN, DT, NB, Random Tree, RF, Rules.Part, SVM, etc. &\checkmark &\checkmark &\checkmark &RF and meta. RandomSubSpace as best classifiers. \\
		\cite{Kumar2019}	&EN	&Twitter &Implement SVM, DT, KNN, NB, NN using Particle Swarm Optimization (PSO) to select optimal features. &\checkmark &\checkmark &  &Improve in accuracy using PSO \\
		\hline
	\end{tabular}
\end{table}

\section{Procedure to compute SPR}\label{Sec:Methodology}

\begin{figure}
	\centering
	\includegraphics[width=0.75\linewidth]{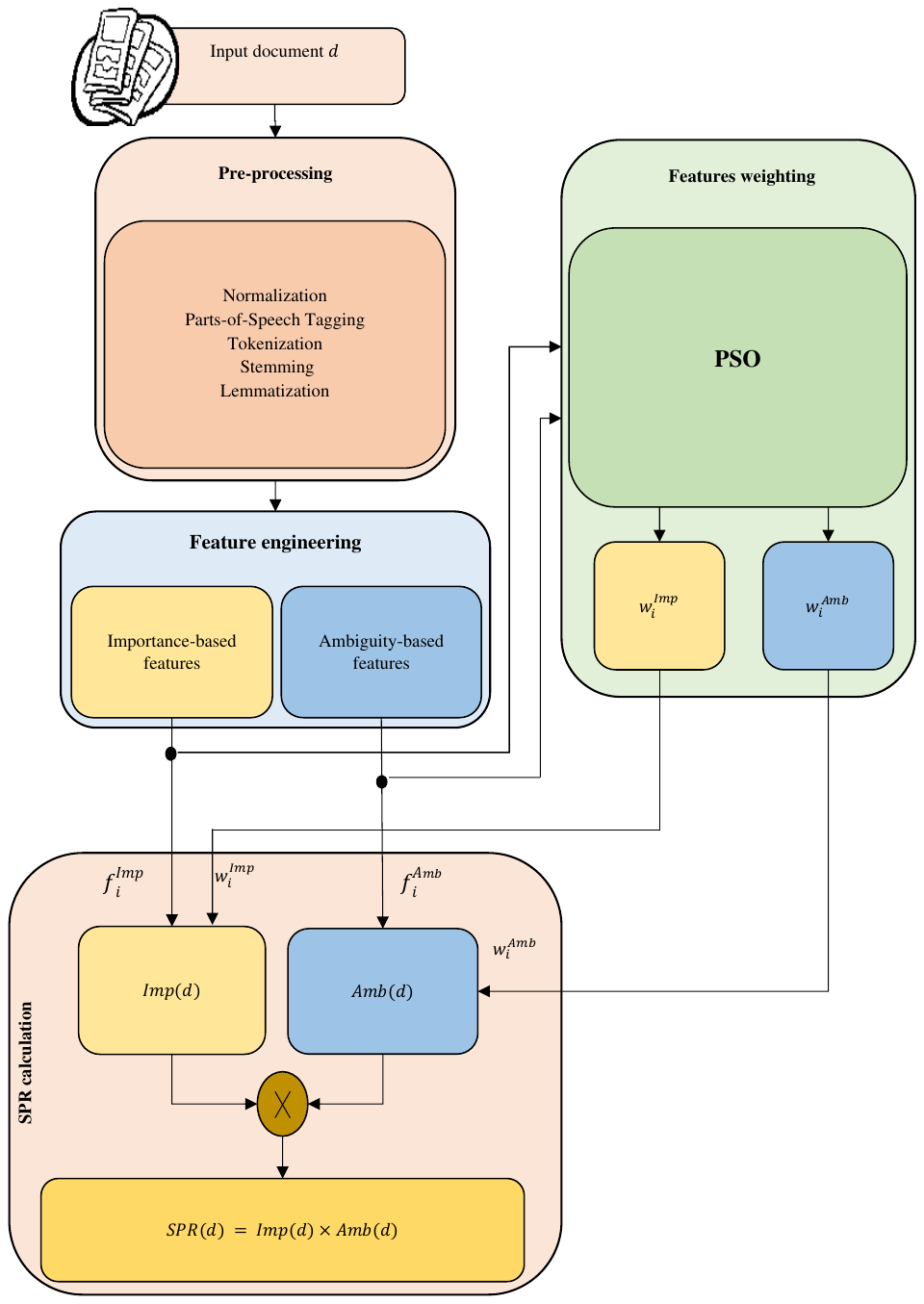}
	\caption{Proposed structure to compute spread power of rumor.}
	\label{Fig1: General Structure}
\end{figure}

In this study, we intend to introduce a new measure called SPR to calculate the spread power of rumor. SPR is a research issue that has not been considered in any of the research on rumor empirically. As mentioned in Section \ref{Sec:theory}, the spread of the rumor depends on the existence of both factors of importance and ambiguity in the rumor. Thereby, in the proposed model, a set of content features in two categories are introduced to compute the importance and ambiguity of a message document with the aim of calculating SPR. The general structure of the proposed model for computing SPR is shown in Figure \ref{Fig1: General Structure}. As described in Figure \ref{Fig1: General Structure}, our model consists of the following steps:

\begin{enumerate}
	\item Pre-processing: Converting the message document into a form that is analyzable for this task.
	\item Feature engineering: Analyzing and extracting the features that indicate the importance and ambiguity of a rumoe document.
	\item Feature weighting: Weighting content features and determining the degree of importance of each feature in predicting classes.
	\item SPR calculation: Computing the Spread Power of Rumor based on two criteria of importance and ambiguity.
\end{enumerate}

\subsection{Pre-processing}\label{Subsec: Text Pre-processing}
The proposed model performs the SPR calculation in the Persian rumors on Twitter and Telegram datasets. These online texts usually contain lots of noise and uninformative parts (such as symbols, special characters). Therefore, five steps of pre-processing, including tokenization, normalization, Part Of Speech (POS) tagging, stemming, and lemmatization are performed on the message documents to bring documents into a form that is usable and analyzable for our task. Pre-processing operations in the Persian language have complexities and challenges that these challenges are addressed by normalization. Some of these challenges include:

\begin{enumerate}
    \item Multiform words: some words may be written in some different forms. For example, \setcode{utf8} \<'مساله'>, \<'مسئله'>, and \<'مسأله'> are all forms of writing the word ‘problem’. It is solved by a spell checker to normalize text into a standard one and unify those words.
    \item Different spacing: some words may be written with space, short space or no space such as: \setcode{utf8} \<'ميگفت'>, \<'مي گفت'>, and \<'مي گفت'> are all forms of writing the word 'was saying'. It is solved by Adding short-spaces between different parts of a word.
    \item Letters with two Unicodes: there are some letters have two Unicodes that one is for Persian and one for Arabic, such as: \setcode{utf8} \<'ی'>, \<'ي'> (i) and \<'و'>, \<'ؤ'> (v). It is solved by Replacing Arabic letters with their Persian equivalent.
\end{enumerate}

\subsection{Feature engineering}\label{Subsec: Feature extraction}
Feature engineering is the process of using domain knowledge to extract features (characteristics, properties, attributes) from raw data \cite{ng2013machine}. We focused on the content of rumor document as an informative source and extracted the content features that increase two influential factors in the spread of rumors, i.e., importance and ambiguity. Because, rumormongers use the power of words in expressing FRs to captivate the audience and gain their trust. The role of an FR is not out of two modes; either it is expressed based on imagination, lies, and slander, or it is published an event that its acceptance depends on the state of the audience's public opinions and its publication time. Thereby, FRs make a sense similar to the truth for the audience so that the audience accepts it and propagates it, even if its validity is doubtful. Hence, FRs are quickly accepted and propagated by audiences without any review of its accuracy.

\begin{figure}
	\centering
	\includegraphics[width=\linewidth]{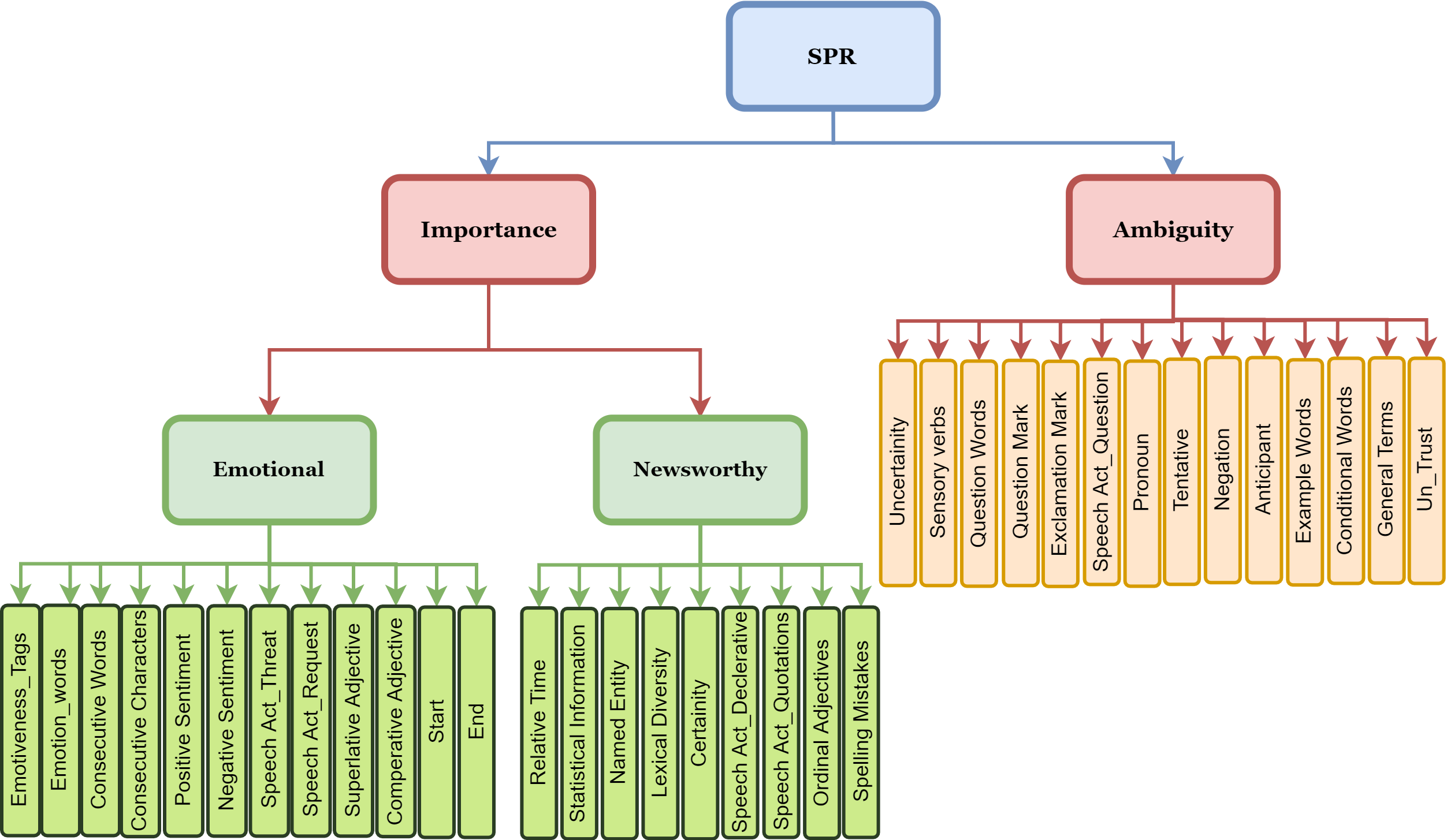}
	\caption{The hierarchical structure of feature engineering for the SPR calculation.}
	\label{Fig:features-chart}
\end{figure}

We introduced 42 features to compute SPR: 28 features to determine the importance of the message document and 14 features to compute the ambiguity of the document. In  Figure \ref{Fig:features-chart} is illustrated the hierarchical structure of content features that are extracted by feature engineering methods for the SPR calculation. Also,  Table \ref{tbl: 1}, \ref{tbl: Ambiguity features}, \ref{tbl: Newsworthy features} show these features along with a brief description of each.

\begin{table}
	\centering
	\caption{A summary of Emotional features along with a brief description of each (The new features are marked with a "*").}
    \label{tbl: 1}
	\begin{tabular}{m{1.2cm} m{2cm} m{12cm}}
	\textbf{Abbr.} &\textbf{Feature} &\textbf{Description}
	\\
	\hline
	\multicolumn{3}{c}{\textbf{Emotional features}}
	\\
	\hline
	ETag	&Emotiveness \cite{Zhou2004}	&The ratio of adjectives plus adverbs to nouns plus verbs.	\\
	\rowcolor{gray!15} Fr	&Fear*	&The ratio of the number of sentences containing fear-based words to the total number of sentences in the document.	\\
	Su	&Surprise*	&The ratio of the number of sentences containing surprise-based words to the total number of sentences in the document.	\\
	\rowcolor{gray!15} Dsg	&Disgust*	&The ratio of the number of sentences containing disgust-based words to the total number of sentences in the document.	\\
	Sad	&Sadness*	&The ratio of the number of sentences containing sadness-based words to the total number of sentences in the text.	\\
	\rowcolor{gray!15} An	&Anger*	&The ratio of the number of sentences containing anger-based words to the total number of sentences in the document.	\\
    Aff	&Affective*	&The ratio of the number of sentences containing affective-based words (Words that cause emotion or feeling, such as, \setcode{utf8} \<'اخطار'> "/"ekhtar"/" Warning")to the total number of sentences in the document.	\\
	\rowcolor{gray!15} MV	&Motion Verbs*	&The ratio of the number of sentences containing motion verbs (such as, jump, dilatory, rotation and so on.) to the total number of sentences in the document.	\\
	CW	&Consecutive Words*	&The ratio of the number of sentences containing consecutive repeated words (such as,\setcode{utf8} \<'توجه توجه'> / "tavajjoh tavajjoh"/"Attention Attention" and so on.) to the total number of sentences in the document.	\\
	\rowcolor{gray!15} CC	&Consecutive Chars*	&The ratio of the number of sentences containing consecutive repeated characters in a word (such as, \setcode{utf8} \<"سلااااااام"> / "salâââââââm"/"helllllloooo") to the total number of sentences in the document.	\\
	PS	&Positive Sentiment \cite{Castillo2011}	&The ratio the number of positive words in the document to the sum of positive and negative words. If the number of positive and negative words is zero, then PS is zero.	\\
	\rowcolor{gray!15} NS	&Negative Sentiment \cite{Castillo2011}	&The ratio of the number of negative words in the document to the sum of positive and negative words. If the number of positive and negative words is zero, then NS is zero.
	\\
	SA\_Thrt	&Speech Act\_Threat \cite{jahanbakhsh2020speech}	&The SA\_Thrt of a document is determined by SA classifier provided by Jahanbakhsh et al. \cite{jahanbakhsh2020speech}. By this SA, we can promise for hurting somebody or doing something if hearer does not do what we want.	\\
	\rowcolor{gray!15} SA\_Req	&Speech Act\_Request \cite{jahanbakhsh2020speech}	&The SA\_Req (Ie, politely asks from somebody to do or stop doing something) of a document is determined by the SA classifier \cite{jahanbakhsh2020speech}.	\\
	Adj\_Sup	&Superlative Adjective*	&The ratio of the number of sentences containing Adj\_Sup (Simple Adjective + Suffixes \setcode{utf8} \<'ترين'> /tarin/ Number + (\<'مين' + 'اٌمين'> /[o]min/) to the total number of sentences in the document.	\\
	\rowcolor{gray!15} Adj\_Cmp	&Comparative Adjective*	&The ratio of the number of sentences containing Adj\_Cmp (Simple Adjective + Suffixes + \setcode{utf8} \<پسوند 'تر '> /tar/)to the total number of sentences in the document.	\\
	Strt	&Start sentence*		&It analyzes whether the first sentence of the document contains emotion-based words. This feature have a boolean value for each document.	\\
	\rowcolor{gray!15} End	&End sentence*	&It analyzes whether the last sentence of the document contains emotion-based words and words associated with the request. This feature have a boolean value for each document.
	\\
	\hline
	\end{tabular}
\end{table}

\begin{table}
	\centering
	\caption{A summary of Newsworthy features along with a brief description of each (The new features are marked with a "*").}
	\label{tbl: Newsworthy features}
	\begin{tabular}{m{1.2cm} m{2cm} m{12cm}}
	\textbf{Abbr.} &\textbf{Feature} &\textbf{Description}
	\\
	\hline
	\multicolumn{3}{c}{\textbf{Newsworthy features}}
	\\
	\hline
	RT	&Relative Time*	&The ratio of the number of sentences containing RT-based words (such as, \setcode{utf8}  \<'امشب'> / "emšab" /) to the total number of sentences in the document.	\\
	 \rowcolor{gray!15} Adj\_Qty	&Quantity Adjective*& The ratio of the number of sentences containing Quantity Adjective (such as,  \setcode{utf8} \<'برخی'> / "barkhi" /"some",  \setcode{utf8} \<'همه'>/"hame"/"all") to the total number of sentences in the document. 	\\
	 ND	&Numerical Digits \cite{Castillo2011}	&The ratio of the number of sentences containing numeral characteres to the total number of sentences in the document.	\\
	\rowcolor{gray!15} NE	&Named Entity \cite{hamidian2019rumor}	&The ratio of the number of sentences containing NE (In three classes, the person's name, the organization, the location) to the total number of sentences in the document.	\\
	LD	&Lexical Diversity \cite{Zhou2004}	&The ratio of vocabulary to the total number of terms in the document \cite{Zhou2004}.	\\
	\rowcolor{gray!15} Cer	&Certainty \cite{Zhou2004}	&The ratio of certainty-based words to the sum of certainty and uncertainty-based words in the document. If certainty and uncertainty word are zero then certainty score is zero.	\\
	SA\_Dec	&SA\_Declarative \cite{jahanbakhsh2020speech}	&The SA\_Dec (Ie, Transfer information to hearer) of a document is determined by the SA classifier \cite{jahanbakhsh2020speech}.	\\
	\rowcolor{gray!15} SA\_Quot	&SA\_Quotations \cite{jahanbakhsh2020speech}	&The SA\_Quot (Ie, speech acts that another person said or wrote before) of a document is determined by the SA classifier \cite{jahanbakhsh2020speech}.	\\
	Adj\_Ord	&Ordinal Adjective*	&The ratio of the number of sentences containing Adj\_Ord (Number + \setcode{utf8} (\<'ام'>, \<'م'>) /om/) to the total number of sentences in the document.	\\
	\rowcolor{gray!15} SM	&Spelling Mistake \cite{Zhou2004}	&The ratio of misspelled words based on typographical errors to total number of words in the document.	\\
	\hline
	\end{tabular}
\end{table}

\begin{table}
	\centering
	\caption{A summary of Ambiguity features along with a brief description of each (The new features are marked with a "*").}
	\label{tbl: Ambiguity features}
	\begin{tabular}{m{1.2cm} m{2cm} m{12cm}}
		\textbf{Abbr.} &\textbf{Feature} &\textbf{Description}
		\\
		\hline
		Ucer	&Uncertainity \cite{Zhou2004}	&The ratio of uncertainty-based words to the sum of certainty and uncertainty-based words in the document. If certainty and uncertainty word are zero then uncertainty score is zero.	\\
		\rowcolor{gray!15} SV	&Sensory Verb \cite{hamidian2019rumor}	&The ratio of the number of sentences containing SV (Such as, \setcode{utf8} \<'شنيدن'> /"šenidan"/hear and \<'ديدن'> /"didan"/see) to the total number of sentences in the document.	\\
		QW	&Question Word \cite{Castillo2011}	&The ratio of the number of sentences containing QW (Such as, what, when, where, and who) to the total number of sentences in the document.	\\
		\rowcolor{gray!15} QM	&Question Mark \cite{Castillo2011}	&The ratio of the number of sentences containing the question mark '?' or multiple question marks "?????" to the total number of sentences of the document.
		\\
		EM	&Exclamation Mark \cite{Castillo2011}	&The ratio of the number of sentences containing the exclamation mark to the total number of sentences of the document.
		\\
		\rowcolor{gray!15} SA\_Ques	&Speech Act\_Question \cite{jahanbakhsh2020speech}	&The SA\_Ques (Such as, usual questions for information or confirmation) of a document is determined by the SA classifier \cite{jahanbakhsh2020speech}.
		\\
		Pro	&Pronoun \cite{Castillo2011}	&The ratio of the number of sentences containing pronoun (A personal pronoun in 1st, 2nd, or 3rd person) to the total number of sentences of the document.
		\\
		\rowcolor{gray!15} Tntv	&Tentative \cite{hamidian2019rumor}	&The ratio of the number of sentences containing tentative adjective (It describes something that is uncertain and unsure) to the total number of sentences of the document.
		\\
		Neg		&Negation \cite{Kwon2017}	&The ratio of the number of sentences containing Negation words (Units of language, including, words (e.g., not, no, never, incredible) and affixes (e.g., -n’t, un-, any-)) to the total number of sentences of the document.
		\\
		\rowcolor{gray!15} Antcpnt	&Anticipation \cite{Kwon2017}	&The ratio of the number of sentences containing Anticipation-based words to the total number of sentences of the document.
		\\
		Adv\_Exm	&Example Words*	&The ratio of the number of sentences containing Example-based words (such as, \setcode{utf8} \<'همچون'>/hamčon/, \<'همانند'>/hamanand/, and  /masalan/, all are meaning "for example") to the total number of sentences of the document.
		\\
		\rowcolor{gray!15} If	&Conditional words*	&The ratio of the number of sentences containing the conditional conjunctions (such as, if) to the total number of sentences of the document.
		\\
		GT	&General Terms \cite{Castillo2011}	&The ratio of the number of sentences containing the general terms (It Refers to a person (or object) as a class of persons or objects) to the total number of sentences of the document.
		\\
		\rowcolor{gray!15} UT	&Un\_Trust*	&The ratio of the number of sentences containing un\_trust words (such as, lack of trust, distrust, and suspicion) to the total number of sentences of the document.
		\\
		\hline
	\end{tabular}
\end{table}

\subsubsection{Computing the importance of a rumor}\label{Subsubsec: Characteristics of an important rumor}
In this section, the importance of a message document is evaluated based on two factors: emotional and its newsworthy.

\begin{itemize}
	\item{\textbf{\textit{Computing the emotional score of a rumor}}}
	
	We have introduced a set of content features in various categories such as adjectives, adverbs, emotion\_words, and so on to determine the emotional score of a text document. The reason for introducing these features is that rumormongers increase the emotional aspect of the message by utilizing power words. Power words are persuasive, descriptive words that trigger an emotional response. They make us feel scared, encouraged, aroused, angry, and so on. The goal of using power words in FRs is to motivate a person to spread a message. In the following, this set of emotional features are described:

	\textbf{f1: Emotiveness (ETag).} Adjectives(Adj), Adverbs(Adv) describe things and modify other words so change our understanding of things.
		
	\begin{equation}\label{eq:2}
	f_{ETag}^{Emo} (d) = \frac{|Adj(d)| + |Adv(d)|}{|Noun(d)| + |Verb(d)|}
	\end{equation}
		
	where, $f_{ETag}^{Emo} (d)$  is the ratio of adjectives $(|Adj(d)|)$ plus adverbs $(|Adv(d)|)$ to nouns $(|Noun(d)|)$ plus verbs $(|Verb(d)|)$ in the document $d$, which is selected as an indication of expressivity of language \cite{Zhou2004}.
		
	\textbf{f2-f8: Word-emotion (WEF).} Rumormongers use the power of word\_emotion to cause fear, concern, and hatred in the audience. In this study, the National Research Council Canada (NRC) \cite{Mohammad2013} emotion lexicon is utilized to obtain the emotional score of words in the content of Persian rumors in eight basic emotions which are anger, fear, anticipation, trust, surprise, sadness, joy, and disgust. \cite{Mohammad2013} also provided versions of the lexicon in over one hundred languages, such as the Persian language. We manually reviewed and corrected each of these words with the help of two linguists. Five categories of these Word-Emotion features (WEF) including, \textit{Fear} (Fr), \textit{Surprise} (Su), \textit{Disgust} (Dsg), \textit{Sadness} (Sad),  \textit{Anger} (An) are evaluated on the input documents. Also, we introduced the affective-based words (such as, \setcode{utf8} \<'دلخراش'> delkharash"/"irritant") can increase its emotional impact. Additionally,  we considered Motion Verbs (MV) as a new feature to review the content of rumors. MVs are categorized into two categories: (1) Transitional (i.e., top to down, down to top, left to right, right to left, and multi-directional). (2) Self-contained motion (such as oscillation, dilatory, rotation, wiggle, wander, rest). For this purpose, we utilized the set of MVs in the Persian language that is collected by \cite{Golfam2014}. They extracted and analyzed 126 MVs from Dadegan site\footnote{www.dadegan.ir} and Persian Language Database\footnote{www.pldb.ihcs.ac.ir} and other written sources. We narrowed down the list of MVs by selecting MVs that often appear in FRs. Score of each of Word\_Emotion Features $WEF = \{Fr,Su,Dsg,Sad,An,Aff,MV\}$ is calculated by formula \ref{eq:3}.
		
	\begin{equation}\label{eq:3}
	\forall j \in WEF    f_{j}^{Emo} (d) = \frac{\sum_{i=1}^{|S(d)|}{|WEF_j(S_i (d))|}}{|S(d)|},
	\end{equation}
		
	where, $|WEF_j(S_i(d))|$ indicates that sentence $S_i$ of the document $d$ contains feature $j$ of the set $WEF$ or not. $|S(d)|$ shows the number of sentences in document $d$.
		
	\textbf{f9-f10: Consecutive Words or Characters (CW \& CC).} Consecutive words or characters within a sentence in every language is syntactically incorrect. However, rumormonger tries to emphasize the main subject of rumor by repeating consecutive words in the FR. For example, \setcode{utf8} \<'توجه توجه'> /Attention Attention, is a CW and words like \setcode{utf8} \<'سلااااااام'>
	/"salâââââââm"/"Helllllllloooo" and \setcode{utf8} \<'هشداااااااار'> /"Hošdââââââr"/"Alaaaaarm" are words containing CC.
	
	\begin{equation}\label{eq:4}
	f_{CW}^{Emo} (d) = \frac{\sum_{i=1}^{|S(d)|}{|CW(S_i (d))|}}{|S(d)|}
	\end{equation}
	
	\begin{equation}\label{eq:5}
	f_{CC}^{Emo} (d) = \frac{\sum_{i=1}^{|S(d)|}{|CC(S_i (d))|}}{|S(d)|}
	\end{equation}
		
	In formulas \ref{eq:4} and \ref{eq:5} are respectively computed the fraction of sentences containing CW and CC to all sentences $|S(d)|$ of the document $d$. $|CW(S_i (d))|$ and $|CC(S_i (d))|$ have a boolean value for each sentence $S_i$ and indicate that sentence $S_i$ of the document $d$ contains CC and CW or not respectively.
		
	\textbf{f11-f12 Sentiment (PS \& NS).} News sentences usually do not convey any sentiment. For example, in the sentence \textit{"The Coaches of the Persepolis and Oil teams of Tehran, Branko and Ali Daei planted a tree seedling in the league organization on the occasion of the arbor day."}, there is no excitement. Nevertheless, rumors contain several characteristic sentiments (e.g., anger) compared to other types of information \cite{Kwon2013}. On the other hand, there is a general lay belief that FRs are dominated by negative sentiment and polarity \cite{Sunstein}. Although rumors contain negative polarity, they are often expressed in positive polarity. The NRC Emotion Lexicon \cite{Mohammad2013} is used to obtain the sentiment score of Persian words in one of the positive (1), negative (-1), or neutral (0) polarities. We utilized the concepts of sentiment polarity of \cite{Zhang2010} and calculated the sentiment score of the rumor document using an NRC lexicon (Formulas \ref{eq:6} and \ref{eq:7}).
		
	\begin{equation}\label{eq:6}
	f_{PS}^{Emo} (d) =
	\begin{cases}
	0  & if |PSntm(d)| = 0 \: \& \: |NSntm(d)| = 0  \\
	\frac{|PSntm(d)|}{|PSntm(d)| + |NSntm(d)|}  & otherwise
	\end{cases}
	\end{equation}
		
	\begin{equation}\label{eq:7}
	f_{NS}^{Emo} (d) = \begin{cases}
	0  & if |PSntm(d)| = 0 \: \& \: |NSntm(d)| = 0  \\
	\frac{|NSntm(d)|}{|PSntm(d)| + |NSntm(d)|}  & otherwise
	\end{cases},
	\end{equation}
		
	where $|PSntm(d)|$ and $|NSntm(d)|$ are the number of positive and negative terms in document $d$.
		
	\textbf{f13-f14: Threat and Request Speech Acts (SA\_Thrt \& SA\_Req).} The importance of a message from the person's point of view increases when the message transmits critical news and motivates fear in the audience. Hence, individuals spread rumors when they feel anxiety or threat. For example, to increase the speed of the release of a post on the social networks, the rumormonger asks audiences to notify the message as soon as possible to his or her relatives and warns the audience that if they do not inform others, a bad event is may happen. We utilized the SA classifier provided by \cite{jahanbakhsh2020speech}. The value of these two features is a value between 0 and 1.

    \textbf{f15-f16: Superlative and Comparative adjectives (Adj\_Sup, Adj\_Cmp).} There is a set of adjectives that their presence in a document can increase the excitement and importance of the subject. We considered two adjectives of superlative (Adj\_Sup, such as, \setcode{utf8} \<'بهترين'> behtarin"/"best", \setcode{utf8} \<'دومین'> dovvomin"/"second" ) and comparative (Adj\_Cmp, such as, \setcode{utf8} \<'ترسناک تر'> tarsnaktar"/"scarier" ).These adjectives upgrade (ie., one thing or person is superior to another) or diminish the main element of rumor. Score of each of features $SCA = \{Adj\_Sup, Adj\_Cmp\}$ is calculated by formula \ref{eq:27}.

    	\begin{equation}\label{eq:27}
	\forall j \in SCA     f_{j}^{Emo} (d) = \frac{\sum_{i=1}^{|S(d)|}{|SCA_j(S_i (d))|}}{|S(d)|}
	\end{equation}
		
	In formula \ref{eq:27}, $|SCA_j(S_i (d))|$ has a boolean value for each sentence $S_i$ and indicate that sentence $S_i$ of the document $d$ contains feature $j$ of the set $SCA$ or not.
		
	\textbf{f17-f18: Start and End of document (Str \& End).} The start and end sentences of a document are very important in conveying its message because the author expresses the main purpose in these sentences. Therefore, it can have a significant effect on attracting the attention of the audience. Thereby, we introduced two new features to analyze the start sentence of the document based on word-emotion and the end sentence based on both word-emotion and words associated with the request. Both these features have a binary value, which indicates whether the start and/or the end sentences of the message document contains word-emotion and/or words associated with the request.

	\item{\textbf{\textit{Computing the newsworthy of a rumor}}}
	The main reason for accepting a message document as credible news by a person is that it is newsworthy. A news item can be defined as "newsworthy information about recent events or happenings, especially as reported by news media."\footnote{\url{https://www.mediacollege.com/journalism/news/newsworthy.html}}. It can be concluded that the content features that increase the publishing power of a text are more in FRs than in TRs. In this study, the set of content features have been introduced to evaluate the newsworthy of the document, including relative time, statistical information, named entity, lexical diversity, certainty, two SAs (including, Declarative and Quotations), ordinal adjective, and spelling mistakes. These factors are detailed below.

	\textbf{\textit{f19: Relative Time (RT).}} If a story happens today, it is news, but when the same thing happened last week, it is no longer interesting. Thereby, the novelty of news is particularly important. For example, journalists write news of the day with past events from a new angle or view daily. Rumormonger also uses RT-based features (such as, "tonight", \<'امروز'> / "emroz" / "today", \<'اخيراً'> / "akhiran" / "recently" and so on) to apparently display new news or tries to pretend that an important event will happen soon. For example, for over three years, the rumor \textit{"Recently, Google has put an Internet voting to change the name of the Persian Gulf"} be released on social networks. So, words with the Adv\_Time tag are extracted as RT and used to calculate the RT score of document $d$ by \ref{eq:9}.
	
	\begin{equation}\label{eq:9}
	f_{RT}^{Nws} (d) = \frac{\sum_{i=1}^{|S(d)|}{|RT(S_i(d))|}}{|S(d)|}
	\end{equation}
	
	In formula \ref{eq:9}, $|RT(S_i(d))|$  indicates whether sentence $S_i$ of the document $d$ contains RT-based words or not.
	
	\textbf{\textit{f20-f21: Statistical Information (SI).}}  SI-based words are divided into two categories: (i) The number of numeral characters, for example, the numeral "56" has two digits: 5 and 6. (ii) Quantity Adjective (Adj\_Qty), words such as \setcode{utf8} \<'برخی'> / "barkhi" / "some, \<'همه'> / "hame" / "all" and so on. \cite{Zhang2015} found that rumors that are short and contain numbers are more likely to be true than those that are long and do not contain any quantitative details. Thereby, we considered two categories of statistical-based features include Quantity Adjective (Adj\_Qty) and Numerical Digits (Num). Score of each of features $SI = \{Num, Adj\_Qty\}$ is calculated by formula \ref{eq:28}.

    	\begin{equation}\label{eq:28}
	\forall j \in SI     f_{j}^{Nws} (d) = \frac{\sum_{i=1}^{|S(d)|}{|SI_j(S_i (d))|}}{|S(d)|}
	\end{equation}
		
	In formula \ref{eq:28}, $|SI_j(S_i (d))|$ has a boolean value for each sentence $S_i$ and indicate that sentence $S_i$ of the document $d$ contains feature $j$ of the set $SI$ or not.
	
	\textbf{\textit{f22: Named Entity (NE).}} Most people follow the topics that are discussed about celebrities. Celebrities as NEs are high-interest items for individuals. They spread related news about famous people based on popularity or disgust. Therefore, rumormongers utilize the names of famous people, scientists, philosophers, organizations, or institutions to increase the newsworthy of the rumor. In this study, NEs are extracted using a Hidden Markov Model (HMM)-based model \cite{Moradi2017} in three classes, the person's name, organization, and location.
	
	\begin{equation}\label{eq:10}
	f_{NE}^{Nws} (d) = \frac{\sum_{i=1}^{|S(d)|}{|NE(S_i(d))|}}{|S(d)|}
	\end{equation}
	
	In formula \ref{eq:10}, $|NE(S_i(d))|$ indicates whether sentence $S_i$ of the document $d$ contain NE phrases or not.
	
	\textbf{\textit{f23: Lexical Diversity (LD).}} Rumormongers try to attract the attention of audiences to the issue of rumor, so repeat important and emotional words on the subject of a rumor. Therefore, using the repetitive words in the document reduces its LD score. Thereby, FRs have a low LD due to the high repetition of tokens.
	
	\begin{equation}\label{eq:11}
	f_{LD}^{Nws}(d) = \frac{|V(d)|}{|T(d)|}
	\end{equation}
	
	Therefore, the ratio of the number of vocabulary $|V(d)|$ to the total number of terms $|T(d)|$ in the document $d$ is calculated as the LD score \cite{Zhou2004}.
	
	\textbf{\textit{f24: Certainty (Cer).}}Rumormongers use certainty-related words to hide their lies about the FR issue and increase the audience's trust in the subject. The certainty score is calculated as a factor influencing the newsworthy of the rumor by formula \ref{eq:12}.
	
	\begin{equation}\label{eq:12}
	f_{Cer}^{Nws} (d) = \begin{cases}
	0  & if |V_{Cer}(d)| = 0 \: \& \: |V_{Ucer}(d)| = 0  \\
	\frac{|V_{Cer}(d))|}{|V_{Ucer}(d)| + |V_{Cer}(d)|}  & otherwise
	\end{cases}
	\end{equation}
	
	In formula \ref{eq:12}, $|V_{Ucer}(d)|$ and $|V_{Cer}(d)|$  are respectively the number of uncertainty-based and certainty-based vocabularies in document $d$.
	
	\textbf{\textit{f25-f26: Declarative and Quotations Speech Acts (SA\_Dec \& SA\_Quot).}} These two types of SA give a formal concept to document. Sometimes, FRs are formally expressed and refer to reliable sources to gain the audience's trust. We determined these SAs in the document using a SA classifier \cite{jahanbakhsh2020speech}. The value of these two features is a value between 0 and 1.

	\textbf{\textit{f27: Ordinal adjectives (Adj\_Ord).}} These Adjectives denote in what order as first, second, third, fourth, and so on. This feature has a boolean value for each sentence, meaning it is checked whether the sentence contains Adj\_Ord or not. The ratio of the number of sentences containing Adj\_Ord to the total number of sentences in the document is calculated by formula \ref{eq:29}.

	\begin{equation}\label{eq:29}
	f_{Adj\_Ord}^{Nws} (d) = \frac{\sum_{i=1}^{|S(d)|}{|Adj\_Ord(S_i(d))|}}{|S(d)|}
	\end{equation}
	
	In formula \ref{eq:29}, $|Adj\_Ord(S_i(d))|$ indicates whether sentence $S_i$ of the document $d$ contain $Adj\_Ord$ or not.

    \textbf{\textit{f28: Spelling Mistake (SM).}} The presence of a misspelling in the text reduces its newsworthy. We utilized Virastyar\footnote{\url{https://virastyar.ir/}} \cite{kashefi2010automatic} as a spell checker to find misspelled words based on typographical errors in the Persian language and calculate Spelling Mistake (SM) to the total number of terms (T) in document $d$ (formula \ref{eq:30}).

	\begin{equation}\label{eq:30}
	f_{SM}^{Nws} (d) = \frac{|SM(d)|}{|T(d)|}
	\end{equation}
\end{itemize}

\subsubsection{Computing the ambiguity of a rumor}\label{Subsubsec: Computing the ambiguity of a rumor}
The essence of rumors is in their ambiguity so that the ambiguity of evidence makes the process of spreading rumors more widely \cite{Nkpa1975}. The ambiguous expression of news challenges the audience. The ambiguity arises when either the news is received in distorted form or the person received contradictory news, and or one cannot understand such news. In the following, a set of ambiguity-based features is introduced that the presence of those features in the document causes ambiguity in the subject.

\textbf{\textit{f29: Uncertainty (Ucer).}} Words that indicate the lack of sureness about someone or something. Rumormonger tries to challenge the audience's mind by creating a sense of uncertainty about the issue. Thereby, a collection of uncertainty-based words in the Persian language is extracted to measure the uncertainty score of the document $d$.
	
\begin{equation}\label{eq:15}
f_{Ucer}^{Amb} (d) = \begin{cases}
0  & if |V_{Cer}(d)| = 0 \: \& \: |V_{Ucer}(d)| = 0  \\
\frac{|V_{Ucer}(d))|}{|V_{Ucer}(d)| + |V_{Cer}(d)|}  & otherwise
\end{cases}
\end{equation}
	
\textbf{\textit{f30: Sensory Verbs (SV).}} These verbs describe one of the five senses of sight, hearing, smell, touch, and taste. For example, \setcode{utf8} \<'شنيدن'> /"šenidan"/hear, \<'حس کردن'> /"hes kardan"/feeling, \<'ديدن'> /"didan"/see, and so on. When a rumormonger creates a rumor, there is a clear sign in the sentence that indicates that he has personally seen or heard what he speaks about it, or it is the result of his reasoning and speculation. These signs are SVs that create evidentiality in rumors. Evidentiality is a grammatical category that its role is to show the source of information. Of course, these verbs appear in cases where the rumormongers want to increase the rumor's credibility, so they use these verbs as a means to emphasize the rumor.
	
\begin{equation}\label{eq:16}
f_{SV}^{Amb} (d) = \frac{\sum_{i=1}^{|S(d)|}{|SV(S_i(d))|}}{|S(d)|}
\end{equation}
	
$|SV(S_i(d))|$ in equation \ref{eq:16} has a boolean value and indicates whether the sentence $S_i$ of the document $d$ contains a sensory verb or not.
	
\textbf{\textit{f31-f33: Question Speech Act and Tokens (SA\_Qes, QW and QM).}} Question Word (QW) is a function word used to ask a question. Therefore, QW, Question Mark (QM), and Speech Act\_Question (SA\_Ques) are considered as factors that raise questions in the mind of the audience, create ambiguity in rumor, and disturb the reader's mind. QM and QW calculate by formulas \ref{eq:17} and \ref{eq:18} respectively.
	
\begin{equation}\label{eq:17}
f_{QW}^{Amb} (d) = \frac{\sum_{i=1}^{|S(d)|}{|QW(S_i(d))|}}{|S(d)|}
\end{equation}
	
\begin{equation}\label{eq:18}
f_{QM}^{Amb} (d) = \frac{\sum_{i=1}^{|S(d)|}{|QM(S_i(d))|}}{|S(d)|}
\end{equation}
	
$|QW(S\_i(d))|$ and $|QM(S\_i(d))|$ separately means that the sentence $S_i$ of document $d$ has at least one question word or question mark or not.
	
\textbf{\textit{f34: Exclamation Mark (EM).}} A punctuation mark is usually used after an interjection or exclamation to indicate strong feelings or high volume (shouting) or to show emphasis. Exclamation mark used for any other purpose, as to draw attention to an obvious mistake, beside the notation of a move considered a good one, (in mathematics) as a symbol of the factorial function.

\textbf{\textit{f35: Pronouns (Pro).}} Rumormongers use less self-reference (first-person singular pronoun), more group-reference (first-person plural pronoun), and other references (third-person pronouns) to create non-immediacy and uncertainty in their rumors.
	
\begin{equation}\label{eq:19}
f_{Pro}^{Amb} (d) = \frac{\sum_{i=1}^{|S(d)|}{|Pro(S_i(d))|}}{|S(d)|}
\end{equation}
	
$|Pro(S_i(d))|$ is the number of sentences containing pronoun (i.e., third-person and first-person plural pronouns).
	
\textbf{\textit{f36: Tentative (Tntv).}} The adjective tentative is used to describe what is unclear. Therefore, rumormongers utilized these types of words to create a sense of hesitation in the audience and engage minds.
	
\begin{equation}\label{eq:20}
f_{Tntv}^{Amb}(d) = \frac{\sum_{i=1}^{|S(d)|}{|Tntv(S_i(d))|}}{|S(d)|}
\end{equation}
	
Thus, a fraction of the sentences $S_i$ of the document $d$ that containing the tentative-based words $|Tntv(S_i(d))|$ is calculated by formula \ref{eq:20}.
	
\textbf{\textit{f37: Negation (Neg).}}  In rumors, the use of negative words refers to two purposes: (1) creating negative emotions, (2) an unusual expression of the news event. In Persian language seven negative prefixes are used to build words with negative or contrastive meaning. These prefixes are: (1)'bi-'/im-(e.g, impolite), (2)un-, in- (e.g., injustice), no-), (3)'zed-'/unti-(e.g., unti-security), (4)'gheir-'/un-(e.g., Unnecessary), (5)'ne-'/'na-'/not(e.g., nemidanam/I do not see, nasalem/unhealthy), (6) hich-/no- (e.g., nobody), (7)la-/without.
	
\begin{equation}\label{eq:21}
f_{Neg}^{Amb} (d) = \frac{\sum_{i=1}^{|S(d)|}{|Neg(S_i(d))|}}{|S(d)|}
\end{equation}
	
In formula \ref{eq:21}, $|Neg(S_i(d))|$ is the number of sentences containing the negative prefixes.

\textbf{\textit{f38: Anticipation (Ancpnt).}} Anticipation-based words are words that (1)coming or acting in advance (for example, clouds anticipant of a storm). (2) Expectant (for example, anticipating: a team anticipant of victory). Many people are interested in predicting many events, so they try to anticipate the most likely problems, but it is impossible to be prepared for each eventuality. The rumormonger also intends to create fear and turmoil in the community by anticipating unpopular events that have not yet happened.

\begin{equation}\label{eq:22}
f_{Ancpnt}^{Amb} (d) = \frac{\sum_{i=1}^{|S(d)|}{|Ancpnt(S_i(d))|}}{|S(d)|}
\end{equation}

$Ancpnt(S_i(d))$ is a binary value that indicate whether the sentences $S_i$ of the document $d$ contains the anticipate-related words or not.

\textbf{\textit{f39: Example Words (EW).}} These words in a sentence can provide more context and help to better understand proper usage. Rumormonger uses these words to generalize the issue and get the audience's attention.

\begin{equation}\label{eq:24}
f_{EW}^{Amb} (d) = \frac{\sum_{i=1}^{|S(d)|}{|EW(S_i(d))|}}{|S(d)|}
\end{equation}

$EW(S_i(d))$ is a binary value that indicate whether the sentences $S_i$ of the document $d$ contains the example\_words or not.

\textbf{\textit{f40: Conditional words (If).}} Conditional conjunctions can be a single word like "if" or several words like "as long as". Rumormonger uses different conditional conjunction to describe the necessary condition for the occurrence of an issue. The use of different conditional conjunction can have a major impact on changing the audience's attitude towards something.

\begin{equation}\label{eq:25}
f_{If}^{Amb} (d) = \frac{\sum_{i=1}^{|S(d)|}{|If(S_i(d))|}}{|S(d)|}
\end{equation}

$If(S_i(d))$ is a binary value that indicate whether the sentences $S_i$ of the document $d$ contains the conditional words or not.

\textbf{\textit{f41: General Terms (GT).}} The general term is the name of a group or a category of a set of things, people, ideas, and the likes. Rumormonger usually uses these terms to discuss an issue as a whole. Examples of general words include furniture, money, equipment, seasoning, and shoes.

\begin{equation}\label{eq:26}
f_{GT}^{Amb} (d) = \frac{\sum_{i=1}^{|S(d)|}{|GT(S_i(d))|}}{|S(d)|}
\end{equation}

$GT(S_i(d))$ is a binary value that indicate whether the sentences $S_i$ of the document $d$ contains the general terms or not.

\textbf{\textit{f42: Un\_Trust (UT)}} The existence of words containing un\_trust in expressing news about famous people or important factors of society causes doubts about the subject in the mind of the audience. Un\_Trust words are words like lack of trust, distrust, suspicion, mistrust, doubt, disbelief, dubiety, wariness, and so on.

\begin{equation}\label{eq:23}
f_{UT}^{Amb} (d) = \frac{\sum_{i=1}^{|S(d)|}{|UT(S_i(d))|}}{|S(d)|}
\end{equation}
	
$UT(S_i(d))$ is a binary value that indicate whether the sentences $S_i$ of the document $d$ containing the UT-based words are calculated.

\subsection{Feature weighting} \label{subsec:Feature weighting}
The different features can have different levels of importance for prediction in classification problems. feature selection and feature weighting approaches \cite{abualigah2017unsupervised} are used to improve the classification of high dimensional data \cite{abualigah2018novel}. In this study, we utilized feature weighting to determine the degree of importance of each feature in predicting two classes FR and TR. The purpose of feature weighting is to determine the degree of importance of each feature in predicting two classes FR and TR. So the weight of each feature will also be effective in calculating SPR. In this step, Particle Swarm Optimization (PSO) \cite{Kennedy2011} is selected among two optimization algorithms including: PSO \cite{Kennedy2011}, Forest Optimization Algorithm (FOA) \cite{Ghaemi2014}, to find optimal weights for each feature. Therefore, high-weight features will be more effective in the classification results. The algorithm of feature weighting as follows:

\begin{algorithm}
	\caption{Feature weighting algorithm}
	\label{Algorithm:feature_weighting}
	\begin{algorithmic}[1]
		\State Feeding algorithm by the extracted features in \ref{Subsec: Feature extraction}
		\State Utilizing the cross-validation method to separate dataset into training and testing set.
		\State Setting up parameters of PSO for each training set, generating randomly all particles’ positions and velocity, setting up the learning parameters, the inertia weight, and the maximum number of iterations.
		\State Updating the velocities of all particles at each iteration.
		\State Training SVM classifier according to particles values.
		\State Calculating the corresponding fitness function for each particle.
		\State Obtaining the best gene weights and best kernel parameters values.
		\State Training SVM classifier with obtained parameters.
		\State Updating the inertia weight and return to step 4.
	\end{algorithmic}
\end{algorithm}

Each of these features has a different impact on measuring the importance and ambiguity scores of a document. Therefore, it is necessary to consider the influence coefficient of each feature in distinguishing FRs from TRs for computing SPR. In the following,the Ambiguity criterion  $(Amb(d))$ and the values of Emotional $(f_{Emo}^{Imp})$ and the Newsworthy $(f_{Nws}^{Imp})$ criteria, which are the two determining factors in calculating the Importance criterion  $(Imp(d))$ are calculated based on the content features introduced in Section \ref{Subsec: Feature extraction} and the weights obtained for each feature by PSO.
\begin{equation}\label{eq:8}
	f_{Emo}^{Imp} (d) = {\frac{\sum_{j = 1}^{|F_{Emo}|}{wf_j \times f_j^{Emo}(d)}}{|F_{Emo}|}}
\end{equation}

In formula \ref{eq:8}, $wf_j$ is weight of feature $f_j^{Emo}$ and $f_j^{Emo}(d)$ is the value of feature $j$ in $F_{Emo} = \{E\_Tag, Fr, Su, D, Sad, An, Aff, MV, PS, NS, CW, CC, SA\_Thrt, SA\_Req, Adj\_Sup, Adj\_Cmp, Strt, End\}$ of document $d$.

The Newsworthy score of document $d$ is computed by formula \ref{eq:13}, where $wf_j$ is weight of feature $f_j^{Nws}$ and $f_j^{Nws}(d)$ is the value of feature $j$ in $F_{Nws} = \{RT, ND, Adj\_Qty, NE, LD, Cer, SA\_dec, SA\_Quot,\\ Adj\_Ord\}$ of document $d$. The presence of Spelling Mistakes (SM) in the text has a negative effect and reduces its newsworthy. Therefore, we subtracted the value of this feature from the sum of other features that increase the newsworthy of the text.

\begin{equation}\label{eq:13}
	f_{Nws}^{Imp} (d) = {\frac{(\sum_{j = 1}^{|F_{Nws}|}{wf_j \times f_j^{Nws} (d)) - (wf_{SM} \times f_{SM}^{Nws} (d))}}{|F_{Nws}|}}
\end{equation}

Also, the ambiguity score of a document is calculated by formula \ref{eq:amb} where, $wf_j$ is weight of feature $f_j^{Amb}$ and $f_j^{Amb}(d)$ is the value of feature $j$ in $F_{Amb} = \{Ucer, SV, QW, QM, EM, SA\_ques, Pro, Tntv, Neg, Antcpnt, \\Adv\_Exm, If, GT, UT\}$ in the document $d$.

\begin{equation}\label{eq:amb}
	Amb(d) = {\frac{\sum_{j = 1}^{|F_{Amb}|}{wf_j \times f_j^{Amb}(d)}}{|F_{Amb}|}}
\end{equation}


\subsection{SPR calculation} \label{subsec: power-calculation}
In this section, SPR score is calculated based on two criteria of Importance and Aambiguity. We defined the Importance of a rumor based on the sum of two factors  Emotional $(f_{Emo}^{Imp})$ and the Newsworthy $(f_{Nws}^{Imp})$ (formula \ref{eq:14}). These two factor are calculated based on content features that are introduced in Section \ref{Subsec: Feature extraction} and weights extracted from the PSO as the coefficient of each feature in Section \ref{subsec:Feature weighting}.

\begin{equation}\label{eq:14}
	Imp(d) = f_{Emo}^{Imp} (d) + f_{Nws}^{Imp} (d)
\end{equation}

\begin{figure}
	\centering
	\includegraphics[width=0.8\linewidth]{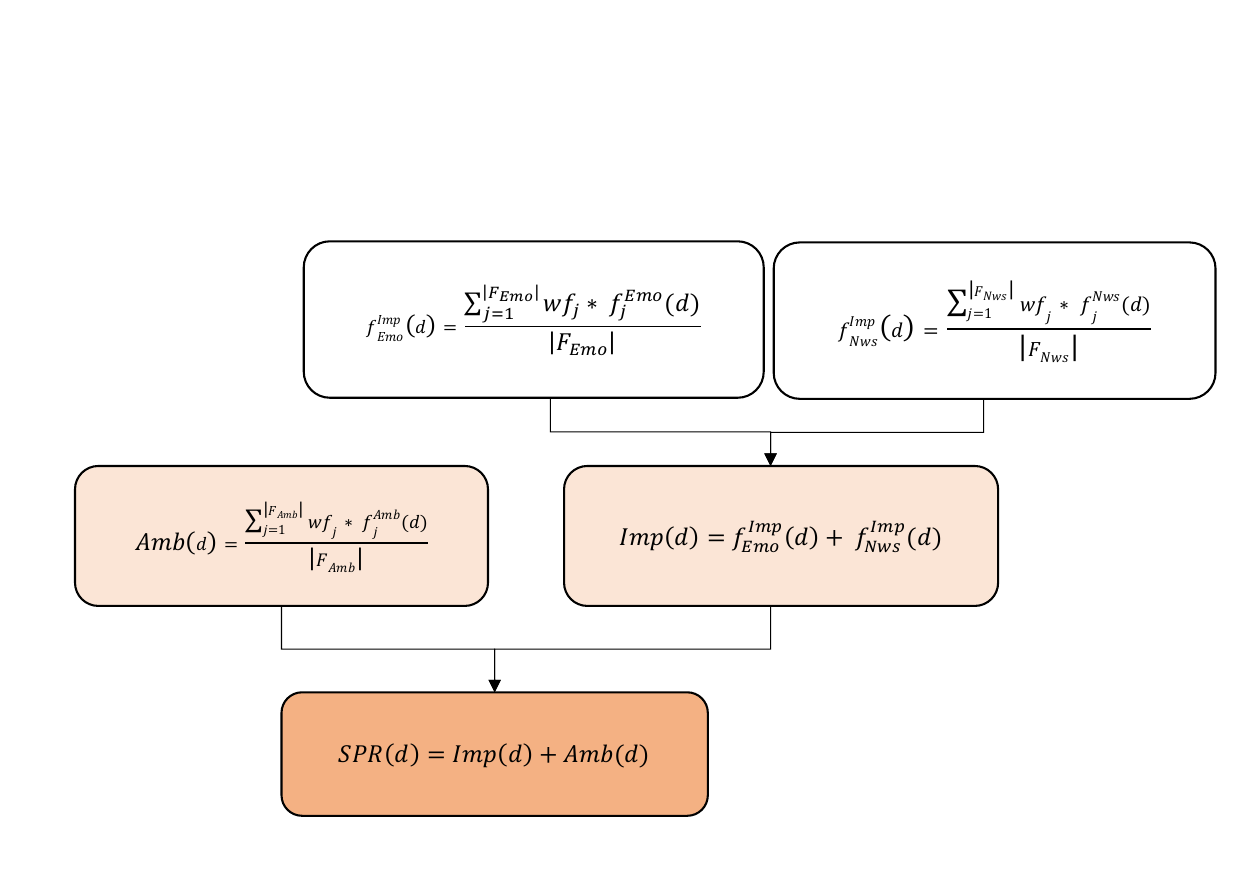}
	\caption{The general procedure of the SPR calculation}
	\label{Fig:SPR-chart}
\end{figure}

According to Allport and Postman's theory, SPR is approximately equal to the multiplication of the importance and ambiguity (formula \ref{eq:allport}) surrounding the rumor. Therefore, in formula \ref{eq:2_4}, the SPR score of document $d$ is calculated based on the multiplication of two scores Importance (eq. \ref{eq:14})  and Ambiguity (eq. \ref{eq:amb}). 

\begin{equation} \label{eq:2_4}
	SPR(d) = Imp(d) \times Amb(d)
\end{equation}

\section{Experiments and Results}\label{Sec:EXPERIMENTS AND RESULTS}
In this section, four experiments are performed on datasets of Twitter and Telegram to evaluate SPR as a new proposed factor and answer to the research questions: (1) Investigating the significant of SPR between the two categories of FRs and TRs. (2) The effect of feature weighting on SPR calculation. (3) Evaluation of presence of SPR on a real world application (Rumor detection). (4) SPR performance on rumor detection. In the experiments presented subsequently, 10-fold cross-validation is used. The experimental evaluation metrics such as Accuracy, Precision, Recall, and F1 measure are used to evaluate the performance of the classifier in identifying rumors in two classes of FR and TR using the Random Forest (RF) classifier. In this section, the experimental details are described.

\subsection{Data collection and dataset of rumors}\label{Subsec: Data Collection and Dataset of Rumors}
This study evaluates SPR on Persian rumors from two different sources: Twitter and Telegram. The details of these two datasets in table \ref{tbl: 3} is described.

\begin{table}
	\centering
	\caption{Distribution of Persian rumors datasets.}
	\label{tbl: 3}
	\begin{tabularx}{\textwidth}{p{1.4cm}p{0.6cm}p{0.6cm}X}
		\textbf{Dataset} & \textbf{FR} & \textbf{TR} & \textbf{Description}	\\
		\hline
		Twitter \cite{Zamani2017}	&783	&783	&Crawling Twitter rumors from two Iranian websites, Gomaneh.com and Shayeaat.ir which publish Persian rumors and annotating by Zamani et al..	\\
		Telegram \cite{jahanbakhsh2020speech}	&882	&882	&Crawling Telegram rumors from three Telegram channels of Iranian websites, Gomaneh.com, Wikihoax.org, and Shayeaat.ir. Also, several Telegram channels (i.e., Fars News Agency, Iranian Students' News Agency (ISNA), Tasnim News Agency, Tabnak, Nasim News Agency (NNA), Mehr News Agency (MNA), Islamic Republic News Agency (IRNA)) has been crawled to extract non-rumors.	\\
	\hline
	\end{tabularx}
\end{table}

\subsubsection{Twitter dataset}\label{Subsubsec: Twitter }
Twitter is a micro-blogging social network service where users can publish and exchange short messages of up to 280 characters long; these messages are called tweets. Accessibility, speed, and ease-of-use have made Twitter a valuable social medium for a variety of purposes that its use is exponentially growing. We utilized the Persian Twitter dataset introduced by \cite{Zamani2017} with the aim of evaluating the SPR score in Twitter rumors.

\subsubsection{Telegram dataset}\label{Subsubsec: Telegram}
Telegram is an instant messaging service. Due to the popularity of Telegram in Iran and the dissemination of messages through it, we considered it to evaluate our work. Jahanbbakhsh et al. \cite{jahanbakhsh2020speech} provided a Persian Telegram dataset for rumor detection. This dataset is a few thousand Persian Telegram posts in two classes TR and FR on various topics, which have been crawled and extracted by using provided API by ComInSys lab\footnote{\href{www.cominsys.ir}{www.cominsys.ir}} of the University of Tabriz and available in \cite{Feizi-Derakhshi2019}.

\subsection{Experiment 1: Investigation of significance of SPR between FR and TR}\label{Subsec: Data Analysis}
In this experiment, the statistical analysis is performed using the T-test on SPR to show its significant difference between the two categoriesFRs and TRs. Also, the result of T-test are investigated on 42  features (Tables \ref{tbl: 1}, \ref{tbl: Newsworthy features}, and \ref{tbl: Ambiguity features}), and the distribution of features in both TR and FR categories is represented by boxplots.

\subsubsection{T-test}\label{Subsubsec: Student t-test}
Since our samples are independent, an independent samples T-test is run on 42 features. An independent samples T-test compares the means and P-value of each feature for two groups FR and TR. NULL hypothesis is rejected if $P < 0.05$. In this study, the null hypothesis is defined as follows:

\begin{center}
\textbf{Null hypothesis:} The spread power of FRs is equal to the TRs.
\end{center}

With the hypothesis that each feature appears with a different frequency in FRs and TRs and can discriminate between them, the P-value is calculated for each feature listed in Tables \ref{tbl: 1}, \ref{tbl: Newsworthy features}, and \ref{tbl: Ambiguity features}. The P-value results $(<=0.05)$ demonstrate that most of these features reveal statistically significant differences between FR and TR documents. Questions 2 \& 3 in Section \ref{Sec:theory} are answered based on p-value results of "Amb" and "Imp". It is indicated that introduced features for computing "Amb" and "Imp" are effective.

\begin{table}
	\caption{The result of t-test for 42 proposed features along with two criteria Ambiguity(Amb) and Importance(Imp) (those values that are less than 0.05 are italicized).}
	\label{tbl: 4}
	\begin{tabular}{lcccccccc}
        \hline
		&\textbf{ETag}  &\textbf{Fr}  &\textbf{Su}	&\textbf{Dsg}	&\textbf{Sad}	&\textbf{An}	&\textbf{Aff}	&\textbf{MV}	\\
		P-value	&0.952	&\textit{0.000}	&\textit{0.000}	&0.094	&\textit{0.000}	&\textit{0.000}	&\textit{0.048}	&\textit{0.000}	\\
		\rowcolor{gray!15}	&\textbf{CW}	&\textbf{CC}	&\textbf{PS}	&\textbf{NS}	&\textbf{SA\_Thrt}  &\textbf{SA\_Req}	&\textbf{Adj\_Sup}	&\textbf{Adj\_Cmp}	\\
		\rowcolor{gray!15} P-value	&\textit{0.000}	&\textit{0.000}	&0.175	&\textit{0.007}	&\textit{0.000}	&\textit{0.000}	&0.305	&\textit{0.000}	\\
		&\textbf{Strt}	&\textbf{End} &\textbf{RT}	&\textbf{Adj\_Qty}	&\textbf{ND}	&\textbf{NE}	&\textbf{LD}	&\textbf{Cer}	\\
		P-value	&\textit{0.000}	&\textit{0.000}	&\textit{0.001}	&\textit{0.000}	&\textit{0.022}	&\textit{0.000}	&\textit{0.000}	&\textit{0.000}	\\
		\rowcolor{gray!15}	&\textbf{SA\_Dec}	&\textbf{SA\_Quot}	&\textbf{Adj\_Ord}	&\textbf{SM}	&\textbf{Ucer}	&\textbf{SV}	&\textbf{QW} &\textbf{QM}	\\
		\rowcolor{gray!15}	P-value	&\textit{0.000}	&\textit{0.033}	&\textit{0.029}	&\textit{0.005}	&0.847	&\textit{0.000}	&\textit{0.000}	&\textit{0.000}	\\
		&\textbf{EM}	&\textbf{SA\_Ques}	&\textbf{Pro}  &\textbf{Tntv}   &\textbf{Neg}	&\textbf{Antcpnt}	&\textbf{Adv Exm}	&\textbf{If}	\\
		P-value	&\textit{0.000}	&\textit{0.004}	&0.228	&\textit{0.002}	&0.515	&\textit{0.000}	 &0.867	&\textit{0.012}	\\
		\rowcolor{gray!15} &\textbf{GT}	&\textbf{UT}	&\textbf{Amb}	&\textbf{Imp}	&	&	&	&	\\
		\rowcolor{gray!15} P-value 	&\textit{0.000}	&\textit{0.000}	&\textit{0.003}	&\textit{0.000}	&	&	&	&	\\
        \hline
	\end{tabular}
\end{table}

Table \ref{tbl: 5} demonstrates the result of the T-test for the SPR feature. Since $p-value = 0.000 \leq 0.05$, the null hypothesis is rejected for SPR, so it shows that there is a significant difference between the spread power of two classes TR and FR. So SPR can be used as a feature in the rumor identification task. This result is the answer to question 4 of Section \ref{Sec:theory} on the ability of SPR in distinguishing between TRs and FRs.

\begin{table}
	\caption{Independent T-test values for SPR}
	\label{tbl: 5}
	\resizebox{\columnwidth}{!}{%
	\begin{tabular}{lm{2.5cm}m{1cm}m{1cm}m{1cm}m{1.25cm}m{1.25cm}m{2cm}m{2cm}m{1.5cm}m{1.5cm}}
		&	&\multicolumn{2}{>{\centering}m{2.5cm}}{\textbf{Levene's Test for Equality of Variances}}	&\multicolumn{7}{c}{\textbf{t-test for Equality of Means}} \\
		\cline{3-11}
		&	&\textbf{F}	&\textbf{Sig}	&\textbf{t}	&\textbf{df}	&\textbf{Sig.(2-tailed)}	&\textbf{Mean Difference}	&\textbf{Std. Error Difference}	&\multicolumn{2}{>{\centering}m{3cm}}{\textbf{95\% Confidence Interval of the Difference}} \\
		&	&	&	&	&	&	&	&	&\textbf{Lower}	&\textbf{Upper} \\
		\cline{3-11}
		\multirow{2}{*}{\rotatebox{90}{SPR}}	&Equal variances assumed	&15.188	&0.000	&4.835	&1233	&\textbf{0.000}	&0.024484	&0.005064	&0.014549	&0.034419	\\
		&Equal variances not assumed	&	&	&4.815	&1131.062	&\textbf{0.000}	&0.024484	&0.005085	&0.014507	&0.034461	\\
	\hline
	\end{tabular}
	}
\end{table}

\subsubsection{Distribution of features in TR and FR}\label{Subsubsec:}
\begin{figure}
	\centering
	\includegraphics[width=\linewidth]{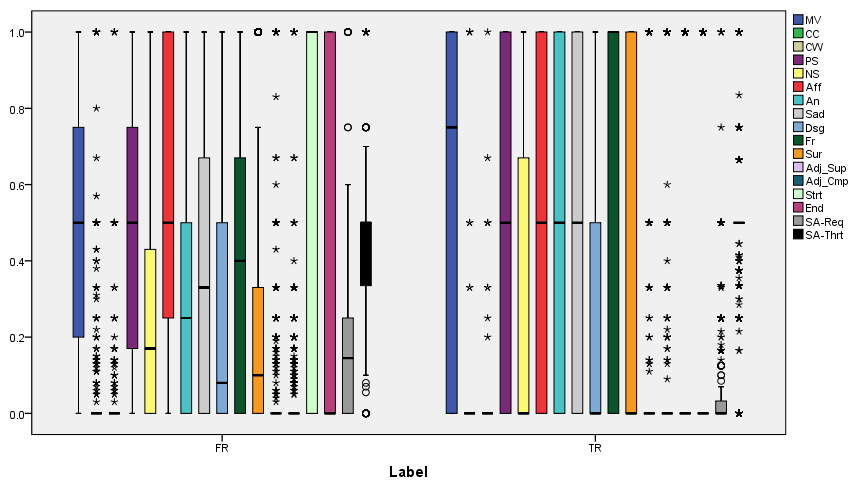}
	\caption{The illustration of the distribution of emotional features by boxplots in two classes of FR (0) and TR (1).}
	\label{Fig2}
\end{figure}

\begin{figure}
	\centering
	\includegraphics[width=\linewidth]{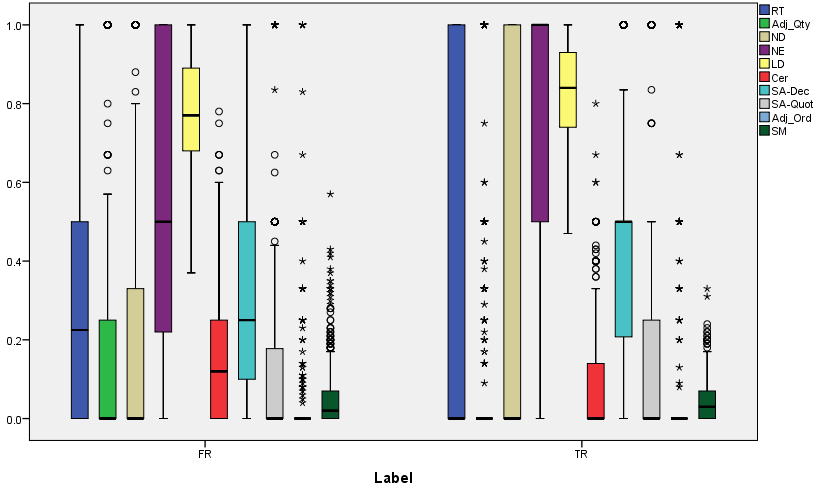}
	\caption{The illustration of the distribution of newsworthy-based features by boxplots in two classes of FR (0) and TR (1).}
	\label{Fig3}
\end{figure}

\begin{figure}
	\centering
	\includegraphics[width=\linewidth]{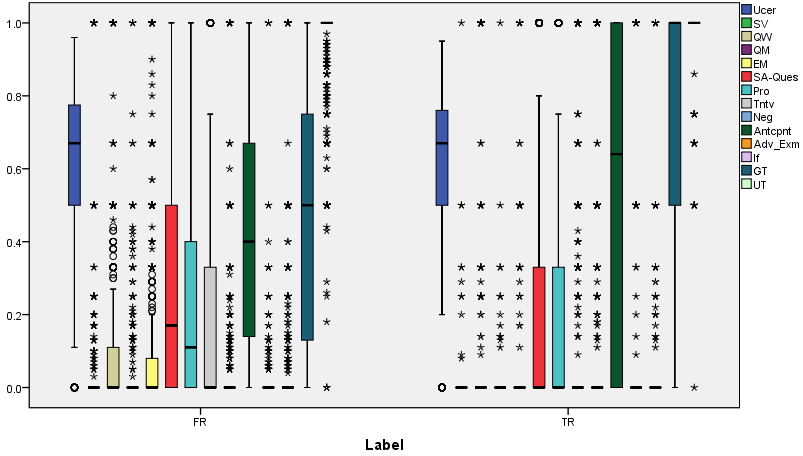}
	\caption{The illustration of the distribution of ambiguity-based features by boxplots in two classes of FR (0) and TR (1).}
	\label{Fig4}
\end{figure}

\begin{figure}
	\centering
	\includegraphics[width=\linewidth]{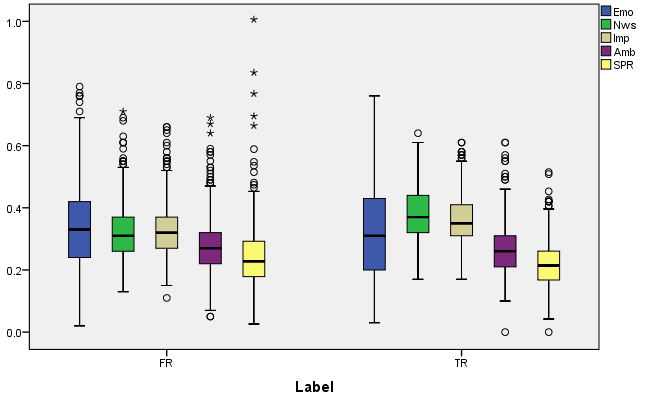}
	\caption{The illustration of the distribution of five features (Emo, Nws, Imp, Amb, and SPR) by box plots in two classes of FR (0) and TR (1).}
	\label{Fig5}
\end{figure}

In this section, the distribution of introduced features for computing the SPR is displayed using the boxplots in two categories TR and FR. The boxplot is a standardized way of displaying the data distribution based on the summary of five numbers: minimum, first quartile, median, third quartile, and maximum. Graphical representation of the distribution of features in three categories "Emotional", "Newsworthy", and "Ambiguity" is shown in Figures \ref{Fig2}, \ref{Fig3} and \ref{Fig4}, respectively. Besides, Figure \ref{Fig5} illustrates the discriminative capacity of five factors of "Emotional", "Newsworthy", "Importance", "Ambiguity", and "SPR" in two classes of FR and TR. As shown in the boxplot diagram (figure \ref{Fig5}), the three features of ambiguity, emotional, and SPR in the FRs have a high distribution than TRs, but newsworthy and importance scores in TRs have a high distribution. Because we considered features such as relative time, statistical information, named entity, lexical diversity, certainty, Declarative SA, Quotations SA, ordinal adjective, and spelling mistakes to compute the newsworthy score which the spelling mistake score has a negative effect on the calculation of news value (Formula \ref{eq:13}), so the presence of spelling mistake in the text reduces the amount of news value of that text. Rumors usually have more spelling mistakes than credible news, because colloquial language is also used to express this type of rumor, which the system identifies colloquial words as misspellings. Spelling mistakes in the text reduce the amount of news value of that text. On the other hand, according to Figure \ref{Fig3}, except for the Adj\_Qty and Cer features, which are highly distributed in rumors, most newsworthy-based features in TRs are highly distributed, which increases the standard value of news value in TR.
	
\subsubsection{The SPR value on Twitter and Telegram datasets} \label{The average of SPR on Twitter and Telegram}
We also investigated the average of the SPR on FRs and TRs in Twitter and Telegram datasets. Therefore, first, the spread power of 1566 FR and TR on Twitter and 1764 FR and TR on Telegram is calculated. Then, the average of SPR is gained for each dataset in two categories of TRs and FRs. The results of the statistical analysis (Table \ref{tbl: 6}) on these two datasets showed that the average of the spread power of FRs is more than TRs in both datasets. Therefore, it can be concluded that the characteristics of a fast-spreading rumor in FRs are more than TRs. As shown in Table \ref{tbl: 6}, the average propagation power in the Twitter data set is lower than the Telegram data set. The reason for this is that the length of tweets is limited, so little content information is extracted from it.
	
\begin{table}
	\centering
	\caption{Comparing the the average values of importance (Imp.), ambiguity (Amb.) and spread power of rumors (SPR) in two categories FRs and TRs on Twitter and Telegram.}
	\label{tbl: 6}
	\begin{tabular}{ccccc}
		\textbf{Dataset}	&\textbf{Category}	&\textbf{Avg. Imp.}	&\textbf{Avg. Amb.}	&\textbf{Avg. SPR}	\\
		\hline
		\multirow{2}{*}{Twitter \cite{Zamani2017}}	&FR	&0.217	&0.137	&0.135	\\
		&TR	&0.274	&0.114	&0.103	\\
		\hdashline
		\multirow{2}{*}{Telegram \cite{jahanbakhsh2020speech}}	&FR	&0.326	&0.274	&0.242	\\
		&TR	&0.361	&0.269	&0.218	\\
    \hline
	\end{tabular}
\end{table}

\begin{table}
	\centering
	\caption{Result of precision, recall and F-score measures of RF classifier based on proposed features to compute the SPR (with and without feature weighting by PSO.}
	\label{tbl: 7}
	\begin{tabular}{ccccc}
		\textbf{Dataset}          & \textbf{Category} & {\begin{tabular}[c]{@{}c@{}}\textbf{Precision}\\ (with/without)\end{tabular}} & {\begin{tabular}[c]{@{}c@{}}\textbf{Recal}\\ (with/without)\end{tabular}} & {\begin{tabular}[c]{@{}c@{}}\textbf{F-measure}\\ (with/without)\end{tabular}}	\\
		\hline
		\multirow{3}{*}{Twitter}	&FR	&0.772 / 0.750	&0.746 / 0.712	&0.759 / 0.730	\\
		&TR	&0.754 / 0.726 &0.780 / 0.763	&0.766 / 0.743	\\
		&Avg.	&0.763 / 0.738 &0.763 / 0.738	&0.763 / 0.737	\\	
		\hdashline
		\multirow{3}{*}{Telegram} &FR	&0.791 / 0.742	&0.814 / 0.781	&0.802 / 0.760	\\
		&TR	&0.825 / 0.768	&0.803 / 0.729	&0.814 / 0.751	\\
		&Avg.	&0.808 / 0.755	&0.808 / 0.755	&0.808 / 0.755	\\
    \hline
	\end{tabular}
\end{table}

\subsection{Experiment 2: The effect of feature weighting on SPR calculation} \label{Subsec: Application of feature weighing in identifying rumors}
In Section \ref{subsec:Feature weighting}, a score is assigned to each feature. These scores indicate the relative importance of each feature when making a prediction. Feature importance scores can provide insight into features extracted from the dataset. The relative scores can highlight which features maybe most or least relevant to the target. In this section, the effect of weighting is evaluated. For this purpose, we have chosen a real application called rumor detection. Therefore, the effect of the feature weighting  on the rumor detection problem is investigated. 

two experiments are performed to show the importance of feature weighting in the rumor detection task: (1) Rumor detection based on the features presented in Tables \ref{tbl: 1}, \ref{tbl: Newsworthy features}, and \ref{tbl: Ambiguity features} regardless of the importance of each feature. (2) Rumor detection based on the same features weighted by PSO. Experimental results in Table \ref{tbl: 7} show the effect of weighting features on the rumor classification process in all three evaluation metrics of Precision (P), Recall (R), and F-measure (F1). Our main goal of weighting features is to make features of high importance more effective in calculating SPR. Therefore, we used the weights extracted by the PSO as the coefficient of each feature in the SPR calculation process. 

\subsection{Experiment 3: Evaluation of presence of SPR on a real world application (Rumor detection)} \label{Subsec: Application of spread power in identifying rumors}
In this section, two experiments are carried out to assess the effect of the SPR score in the rumors classification. This experiment aims to answer Question 4 in Section \ref{Sec:Problem}. For this purpose, two experiments are carried out to assess the effect of the SPR score in the rumors classification. In the first experiment, the classification of two classes of FR and TR is performed based on the set of content-based features. These features include a set of features used in previous studies and a set of new features proposed in this study (Tables \ref{tbl: 1}, \ref{tbl: Newsworthy features}, and \ref{tbl: Ambiguity features}). In the second experiment, the SPR factor is added to these feature sets, and the process of rumors classification is done using the RF classifier.

Table \ref{tbl: 8} shows the result of the evaluation metrics of Precision (P), Recall (R), and F-measure (F1) to evaluate the SPR and its impact on rumors detection. As shown in Table \ref{tbl: 8}, the SPR -as a new feature- has been effective in classifying rumors, and the F-measure has been improved from 0.762 to 0.828.
	
\begin{table}
	\centering
	\caption{The effect of SPR in rumor detection on Telegram dataset using RF classifier.}
	\label{tbl: 8}
	\begin{tabular}{cccccc}
		&\textbf{TP Rate}	&\textbf{FP Rate}	&\textbf{P}	&\textbf{R}	&\textbf{F1}	\\
		\hline
		\multicolumn{6}{c}{\textbf{(1) Content features}}	\\
		FR	&0.753	&0.230	&0.766	&0.753	&0.759	\\
		TR	&0.770	&0.247	&0.757	&0.770	&0.764	\\
		Avg.	&0.762	&0.238	&0.762	&0.762	&0.762	\\
		\multicolumn{6}{c}{\textbf{(2) Content features + SPR}} \\
		FR	&0.802	&0.145	&0.802	&0.846	&0.824	\\
		TR	&0.855	&0.198	&0.855	&0.812	&0.833	\\
		Avg.	&0.828	&0.172	&0.828	&0.829	&\textbf{0.828}	\\
    	\hline
	\end{tabular}
\end{table}

\subsection{Experiment 4: SPR performance on rumor detection} \label{Subsec: SPR}
The performance of the SPR criteria in rumor detection process is compared with the reported results of some existing techniques in the literature on the Persian language (\cite{Zamani2017} and \cite{jahanbakhsh2020speech}) on two datasets (Twitter and Telegram datasets). The proposed method of \cite{Zamani2017} is based on three sets of content, user, and structural features. 

Table \ref{tbl: 9} shows the results of the comparison of our work, \cite{Zamani2017}, and \cite{jahanbakhsh2020speech} on both available datasets (Table \ref{tbl: 3}). The average F-measure of our model to recognize Twitter and Telegram rumors is 0.764 and 0.828, respectively. These results are satisfactory compared to both \cite{Zamani2017} and \cite{jahanbakhsh2020speech} works. Because \cite{Zamani2017} addressed rumor detection on Persian Twitter by analyzing two categories of rumor features: Structural and Content-based features. Their experiments yielded about 70\% precision only based on structural (user graph) features and more than 80\% based on both categories of features on the Twitter dataset. They did not report the experimental result based on content features. Thereby, we re-implemented their work based solely on content features (i.e., about 50000 frequent Twitter unigrams). In this experiment, the average F-measure to recognize Twitter and Telegram rumors is 0.514 (i.e., about 50\% precision) and 0.674, respectively. Hence, we achieved a satisfactory result only by focusing on textual features. Because \cite {Zamani2017} in addition to Unigrams used a set of structural features to categorize rumors. But, our work is only based on content information of rumors, and no other information at the user level and the propagation network is considered. However, our work in detecting rumors with the average F-measure criterion of 0.764 is more satisfactory than \cite{Zamani2017}. On the other hand, our work as a content-based method can detect rumors early, but this feature does not exist in \cite{Zamani2017}'s work, which is based on time-dependent characteristics. Since there is insufficient information about users and the structure of rumors in the early hours of rumor propagation, defining and using effective content features such as SPR can help distinguish rumors from non-rumors.

\cite{jahanbakhsh2020speech} used a set of content features (such as negative and positive sentiment, negation, uncertainty, certainty-related words, lexical diversity, pronoun, depth of dependency tree, word and sentence length, punctuation, number of words, sentences, adjective, adverb, verb, and speech act) to detect Persian Telegram rumors. We also evaluated this work on Twitter dataset. As Table \ref{tbl: 9} shows, we were able to achieve better results in rumor detection by introducing more effective and newer content features such as SPR in comparison with \cite{jahanbakhsh2020speech}. According to the experimental results, the average F-measure to detect Twitter rumors from 0.732 to 0.764 and Telegram rumors from 0.791 to 0.828 improved.


\begin{table}
	\centering
	\caption{Comparison of the proposed method with previous methods to detect Persian rumors based on content-based features analysis.}
	\label{tbl: 9}
	\begin{tabular}{c|cccc|ccc}
		\multirow{2}{*}{\textbf{Method}}	&	& \multicolumn{3}{c}{\textbf{Twitter}}	& \multicolumn{3}{|c}{\textbf{Telegram}}	\\
		& 	& TR	& FR	& Avg	& TR	& FR	& Avg	\\
		\hline
		\multirow{3}{*}{\cite{Zamani2017}}	& Pr	& 0.568	& {0.928}	&0.753	&0.628	&{0.951}	&0.787 	\\
		&Re	&{0.987}	&0.194	&0.587	&{0.981}	&0.412	&0.658	\\
		&F1	&0.721	& 0.320	&0.514	&0.765	&0.575	&0.674	\\
		\hline
		\multirow{3}{*}{\cite{jahanbakhsh2020speech}}	&Pr	&{0.760}	&0.705	&0.734	&0.774	&0.810	&0.792	\\
		&Re	&0.710	&{0.755}	&0.732 	&0.823	&0.760	&0.791	\\
		&F1	&0.734	&0.729	&0.732 	&0.798	&0.784	&0.791	\\
		\hline
		\multirow{3}{*}{Content features + SPR}	&Pr	&0.736	&0.792	&0.766	&0.855	&0.802	&0.828	\\
		&Re	&0.780	&0.750	&0.764	&0.812	&0.846	&0.829	\\
		&F1	&\textbf{0.757}	&\textbf{0.771}	&\textbf{0.764}	&\textbf{0.833}	&\textbf{0.824}	&\textbf{0.828}	\\
	\hline
	\end{tabular}
\end{table}
	
\section{Discussion and Conclusion}\label{Sec:Conclusion}
The content power of the message has a direct effect on the rapid spread of rumors on social networks. The first influential factor in spreading a rumor is its textual content. So it is the power of words that can affect the audience. Accordingly, rumormongers use the power of words to express the rumor and increase its spread power to gain the audience's attention and trust. Hence, we focused on the content characteristics of the source rumor to calculate the Spread Power of Rumor (SPR) as a time-independent measure (i.e., this criterion can be calculated in the early hours of rumor propagation).

The purpose of this paper was to provide a mathematical formula for calculating the Spread Power of Rumor (SPR) for the first time. Determining the spread power of information available on online media is an unaddressed and new task in rumors analysis. The importance and ambiguity are the two main determining factors in creating and spreading of rumors. Therefore, a set of content-based features is engineered in two categories (i.e., importance and ambiguity).

We performed experiments to evaluate SPR. In the first experiment, SPR is investigated based on statistical analysis using the T-test. The T-test' results indicated that SPR reveals statistically significant differences between FR and TR documents. Thereby, our hypothesize about the SPR ability in distinguishing False Rumor (FR) from True Rumor (TR) is confirmed based on achived results. Next, in the second experiment, we showed the positive effect of the weighting of features by PSO and its efficiency in the SPR calculation process. Then in third experiment, in order to perusing the ability of SPR in rumor detection, we have developed a rumor detection system that performs rumor detection through two separate methods; during first method, a set of content features are considered for rumor detection, meanwhile the SPR criterion is also added to this feature set in second method. Experimental results show that the F-measure improved from 0.762 to 0.828. Finally, in the fourth experiment, the performance of the SPR was evaluated in comparison with other related work performed in the field of rumor detection.

Two rumor datasets (Twitter and Telegram) in Persian language are used to evaluate the SPR score. The SPR calculation in Telegram dataset led to the best results and the most accurate estimates compared to the Twitter dataset. Because, Telegram posts are longer than Twitter tweets. So more content information is extracted, and the SPR score is calculated more accurately. Finally, we conclude that SPR as a mathematical formula which is introduced for the first time, can be helpful and effective in early rumor detection. Also, since SPR calculation is independent of time-based features, it can be efficient in detecting rumors in the early hours of propagation.

\section{Future works} \label{Sec:future_work}
Finally, we introduced several open issues that outline promising directions for future research on other applications of SPR.

\begin{itemize}
	\item \textbf{SPR calculation on other language.} We believe that the proposed method is applicable for other languages, because SPR calculation is inherently independent of language. It is based on Allport and Postman's theory and their theory was made by analyzing the psychological aspects of English rumors. Therefore, we intended to extend SPR calculation to other languages.
	\item \textbf{The spread power of advertising messages.} As another application, we can provide more spreadable advertising messages by calculating their spread power. In the other word, SPR would be used in the evaluation of the spread power of advertising messages before their propagation. Because by defining a threshold for the spread power and according to the evaluation results, if the criterion of the spread power obtained for an advertising message was less than the threshold, the advertising message can be strengthened by providing stronger messages for publication. A strong advertising message can make the customer more aware of the brand and increase revenue while reducing the cost of attracting customers by sending a good advertising message.
	\item \textbf{Spread power for detecting hot topics.} As another application, spread power can be used in hot topic detection methods. The task of hot topic detection is to find topics that are frequent during a short period of time. So, the spread power can be used as a feature in any topic detection algorithms to find (powerful) topics.
	\item \textbf{Rumor verification based on SPR.} The SPR can be used as a basic module for a rumor detection system. In the other word, the input of rumor detection system is the raw documents about an event that the SPR score is calculated for each document. The output of the system is a set of documents with a high SPR score that these documents would be considered as the input to the verification system. Therefore, by analyzing the spread power of rumors and retrieving posts with high SPR score, preliminary step can be taken to identify the rumors. In this case SPR acts like a filter for verification system and save verification system’s time.
\end{itemize}

\section*{Acknowledgements}
This project is supported by a research grant of the University of Tabriz (number S/806).

\section*{Declarations}
\subsection*{Data Availability}
Data sharing is not applicable to this article as no new data were created or analyzed in this study.
\subsection*{Competing interests}
The authors declare that they have no conflict of interest.
\subsection*{Funding}
This project is supported by a research grant of the University of Tabriz (number S/806).


\bibliography{mybibfile}

\end{document}